\documentclass[journal]{IEEEtran}

\usepackage[pdftex]{color,graphicx}
\usepackage{amssymb}
\usepackage{amsmath}

\newtheorem{claim}{Claim}

\newtheorem{example}{Example}

\renewcommand{\QED}{\QEDopen}

\begin{document}

% paper title

\title{Signal Codes}

\author{Ofir~Shalvi,
				Naftali~Sommer,~\IEEEmembership{Senior Member,~IEEE,}
        and~Meir~Feder,~\IEEEmembership{Fellow,~IEEE} 
\thanks{N.~Sommer and O.~Shalvi are with the Department of Electrical Engineering-Systems, Tel-Aviv University, Tel-Aviv, Israel, and with Anobit Technologies, Herzlia, Israel. }
\thanks{M.~Feder is with the Department of Electrical Engineering-Systems, Tel-Aviv University, Tel-Aviv, Israel (e-mail: meir@eng.tau.ac.il). }
\thanks{The material in this paper was presented in part at the 2003 IEEE Information Theory Workshop (ITW2003), Paris, France, April 2003.}
         }

\maketitle

\begin{abstract}
Motivated by signal processing, we present a new class of channel codes, called signal codes, for continuous-alphabet channels. Signal codes are lattice codes whose encoding is done by convolving an integer information sequence with a fixed filter pattern. Decoding is based on the bidirectional sequential stack decoder, which can be implemented efficiently using the heap data structure.
Error analysis and simulation results indicate that signal codes can achieve low error rate at approximately 1dB from channel capacity.
\end{abstract}

\section{Introduction}
In this paper we present ``signal codes'', a new approach to channel coding for bandwidth-limited, continuous-alphabet channels such as the band-limited additive white Gaussian noise (AWGN) channel. 
The common approach to signaling for reliable communication over continuous-alphabet channels is based on incorporating coding and modulation, as in trellis coding, to generate points (codewords) in the signal space that belong to a subset (codebook) of the set of all possible modulated values. From a geometric point of view, the transmitted codewords can be considered as constellation points of some high dimensional constellation. 
In many cases these coding and modulation techniques are based on finite-alphabet codes. The set of codewords form a sub-constellation of a denser constellation, where the sub-constellation points satisfy additional constraints induced by the finite-alphabet codewords. 

The approach suggested in this paper is somewhat different, as the high dimensional, high coding gain constellation is designed directly in the Euclidean space, without the help of a finite-alphabet code. We begin with an uncoded signal whose values are drawn from a conventional PAM/QAM constellation. The uncoded signal then passes via a \emph{properly chosen} linear filter that improves the distance spectrum between the different possible signals. %In the signal space, the filtered signal points lie on a lattice. 
To preserve power, a shaping operation projects the filtered signal points into a constrained shaping domain such that the power does not increase, where the shaping operation is based on known pre-coding algorithms. %The filtered pre-coded signals are the codewords of our proposed signal codes.

Linear filtering has been employed in the context of coding in Partial Response Signaling (PRS) and in Faster Than Nyquist (FTN) signaling. See \cite{Rusek_PHD} for an overview of these techniques. Both techniques obtain bandwidth efficiency by introducing a certain amount of
intentional inter symbol interference (ISI). In PRS, the purpose of the ISI is narrowing the power spectrum of the transmitted signal without degrading error probability. In FTN, the purpose is increased data rate. This is done by using a signaling rate which is above the Nyquist rate of the channel, and handling the unavoidable ISI at the receiver. Signal codes differ than these two techniques, as neither the signaling rate, nor the bandwidth are affected: encoding simply transforms discrete valued symbols to continuous valued symbols in a way that improves the error probability without changing the power spectrum of the signal. %Signal codes can be regarded as an n-dimensional constellation design, sone directly at the Euclidean space. %Signal codes utilize pre-coding, that was previously suggested for equalization purposes, i.e., to cancel the effect of inter-symbol-interference (ISI) in the channel. 
%a purposely-induced Inter-symbol interference (ISI) is introduced to generate coding gain, and use pre-coding to preserve or to shape the input to the desired power. 

In fact, signal codes are a special class of lattice codes. In a lattice code, every codeword is of the form $\underline{c}=\boldsymbol{G}\underline{b}$, where $\boldsymbol{G}$, the generator matrix of the lattice, is a real matrix with independent columns and $\underline{b}$ is a vector of integers. For signal codes, the lattice generator matrix has a Toeplitz form. Lattice codes are known to be capable of achieving the AWGN channel capacity (\cite{Debuda1} -- \cite{Zamir_Erez}), and can be interpreted as the Euclidean space analogue of linear binary codes. In this regard, signal codes can be interpreted as the Euclidean space analogue of binary convolutional codes. Note that another family of practical, high coding gain lattice codes are the recently-introduced Low-Density Lattice Codes (LDLC) \cite{LDLC}, which are defined as lattice codes whose generator matrix has a sparse inverse. In \cite{LDLC}, these codes were shown to work as close as 0.6dB to channel capacity with block length of 100,000 symbols. LDLC can be regarded as the Euclidean space analogue of binary Low-Density, Parity-Check (LDPC) codes.

Decoding of signal codes is an equalization problem, and can be done by the Maximum Likelihood Sequence Detector (MLSD) equalization algorithm \cite{Forney_MLSE}. However, the computational complexity of this algorithm is exponential in the number of states and becomes prohibitively large for signal codes with high coding gain. 
We will show that signal codes can approach the AWGN channel cutoff rate with simple sequential decoders, and can also achieve low error rates at approximately 1dB from the AWGN channel capacity, using more elaborate bidirectional stack sequential decoders, whose efficient implementation is based on the heap data structure.

The outline of this paper is as follows. First, signal codes
are defined and presented in Section \ref{signal_code_def}. Then, several shaping algorithms that can be combined with the encoding operation of signal codes are presented in Section \ref{shaping}. Error spectrum analysis and methods to choose the parameters of the code are described in Section \ref{err_spect}, followed by a description of computationally efficient decoders in Section \ref{decoders}. Then, some extensions to the basic signal coding scheme are discussed in Section \ref{extensions}. Simulation results are finally presented in Section \ref{sim_res}.

%\section{Preliminary Results - Signal Codes}
\section{Definition of Signal Codes}  \label{signal_code_def}

%Finally, instead of using arbitrary integers as the information in a lattice code, it is sometimes beneficial to use odd integers. 
We shall first define Pulse Amplitude Modulation (PAM) and Quadrature Amplitude Modulation (QAM) constellations as follows. An $M$-PAM constellation is defined as the set $\{-(M-1), -(M-3),...,-3, -1, 1, 3,..., M-3, M-1\}$. An $M^2$-QAM constellation is defined as the set of complex numbers whose real and imaginary parts belong to an $M$-PAM constellation. A PAM symbol is an integer that belongs to an $M$-PAM constellation, where a QAM symbol is a complex integer that belongs to an $M^2$-QAM constellation. It can be easily seen that the average energy of an $M$-PAM constellation is $(M^2-1)/3$, where the average energy of $M^2$-QAM constellation is $2(M^2-1)/3$.

The motivation for using signal codes comes from considering the effect of linear filtering on the minimum distance of a QAM symbol sequence. %, which is summarized in the  following claim.
%\begin{claim} \label{dist_claim}
Let $\{a_n\}$, $n=0,1,...N-1$ be a random sequence of zero-mean, independent, identically distributed (i.i.d.) QAM symbols. Suppose that $\{a_n\}$ is filtered with a monic causal filter with transfer function $F(z) = 1 +  \sum_{l=1}^L f_l z^{-l}$, %as in (\ref{filt_op}), 
yielding the sequence $\{x_n\}$, $n=0,1,...N+L-1$.
Denote by $d_a^2$ the minimum squared Euclidean distance between two possible $\{a_n\}$ sequences, %Obviously, $d_a^2$ is simply the minimum distance of the input QAM constellation which is $d_a^2=2^2=4$. 
%Denote 
and by $d_x^2$ the minimum squared Euclidean distance between two possible $\{x_n\}$ sequences. The minimum squared Euclidean distances are then related by:
\begin{align} \label{dist_claim}
1 \leq   \frac{d_x^2}{d_a^2}  \leq  \frac{E\{|x_n|^2\}}{E\{|a_n|^2\}}
\end{align}
%\end{claim}
%\begin{proof}
In order to see it, we shall first show that $1 \leq   d_x^2/d_a^2$. Let $x_1(n)$ and $x_2(n)$ be two filtered sequences whose relative distance is the minimum distance $d_x^2$, and let $a_1(n)$ and $a_2(n)$ be the corresponding input sequences. Let $m$ be the smallest index for which $a_1(m) \neq a_2(m)$. Since $F(z)$ is monic and causal, $x_1(m)-x_2(m) = a_1(m)-a_2(m)$, and thus:
\begin{align}
d_x^2=  \sum_n|x_1(n)-x_2(n)|^2 \geq |x_1(m)-x_2(m)|^2 =
\end{align}
\begin{align*}
=|a_1(m)- a_2(m)|^2 \geq d_a^2.
\end{align*}
Turning to the second inequality, let $a_1(n)$ and $a_2(n)$ be two input sequences such that $a_1(n)=a_2(n)+d_a \delta(n)$, where $\delta(n)$ is a Kronecker delta function. Then, the corresponding filter outputs $x_1(n)$ and $x_2(n)$ satisfy $x_1(n)-x_2(n)=d_a f(n)$, so
$d_x^2 \leq  \sum_n|x_1(n)-x_2(n)|^2 =d_a^2 \sum_n |f(n)|^2$. 
On the other hand, due to the i.i.d. assumption on $a(n)$, we have 
$E\{|x_n|^2\}= E\{|a_n|^2\} \sum_n |f(n)|^2$, and the inequality follows.

%\end{proof}

As a consequence, monic linear filtering always improves the minimum distance of an uncoded i.i.d. QAM sequence, but never enough to justify the power increase due to the filtering operation. Therefore, as long as we solve the power increase problem, we have found a way to generate sequences with improved minimum distance, which is a desirable property for coding.  This leads to the definition of signal codes.
%\begin{definition}[signal codes] \label{signal_def}
In signal codes, a sequence of QAM symbols $a_n$
%$\underline{\boldsymbol{x}}$ is generated directly at the $N$-dimensional
%Euclidean space as a linear transformation of a corresponding
%integer message vector $\underline{\boldsymbol{a}}$. This vector, assumed $N-L$ dimensional,
is encoded by %, i.e.,
%$\underline{x}=\boldsymbol{G}\underline{b}$.
convolving its elements with a fixed monic minimum phase filter with transfer function  $F(z) = 1 +  \sum_{l=1}^L f_l z^{-l}$:
\begin{align} \label{filt_op}
x_n = a_n + \sum_{l=1}^L f_l a_{n-l}										
\end{align}
for $n=0,1,...,N+L-1$, where $a_n$ is assumed zero outside the range $0$ to $N-1$. In order to solve the energy increase problem, the $a_n$ sequence has to be modified prior to the filtering operation. This modification will be discussed later.
%\end{definition}

We shall now show that a signal code is a lattice code.
An $n$ dimensional lattice in $\mathbb{R}^m$ is defined as the set
of all linear combinations of a given basis of $n$ linearly
independent vectors in $\mathbb{R}^m$ with integer coefficients. An $n$ dimensional complex lattice in $\mathbb{C}^m$ is similarly defined as the set
of all linear combinations of a given basis of $n$ linearly
independent vectors in $\mathbb{C}^m$ with complex integer coefficients.
The matrix $G$, whose columns are the basis vectors, is called a
generator matrix of the lattice.
A lattice
code of dimension $n$ is defined by a (possibly shifted) lattice
$\boldsymbol{G}$ and a shaping region $B$, where the codewords
are all the lattice points that lie within the shaping region $B$.

According to the above definition, a signal code is a lattice code with the following $(N+L) \times N$ generator matrix:
\begin{align} \label{G_def}
\mathbf{G} =
\left(\begin{array}{ccccccc}
1 & 0 & 0 & \cdots & 0 & 0 & 0 \\
f_1 & 1 & 0 & \cdots & 0 & 0 & 0\\
f_2 & f_1 & 1 & \cdots & 0 & 0 & 0\\
\vdots & \vdots & \vdots & \vdots & \vdots & \vdots & \vdots\\
f_L & f_{L-1} & f_{L-2} & \cdots & 0 & 0 & 0\\
0 & f_L & f_{L-1} & \cdots & 0 & 0 & 0\\
0 & 0 & f_L & \cdots & 0 & 0 & 0\\
\vdots & \vdots & \vdots & \vdots & \vdots & \vdots & \vdots\\
0 & 0 & 0 & \cdots & 1 & 0 & 0\\
0 & 0 & 0 & \cdots & f_1 & 1 & 0\\
0 & 0 & 0 & \cdots & f_2 & f_1 & 1\\
\vdots & \vdots & \vdots & \vdots & \vdots & \vdots & \vdots\\
0 & 0 & 0 & \cdots & f_L & f_{L-1} & f_{L-2}\\
0 & 0 & 0 & \cdots & 0 & f_L & f_{L-1}\\
0 & 0 & 0 & \cdots & 0 & 0 & f_L
\end{array} \right)
\end{align}
%\end{displaymath}
where the encoding operation is equivalent to $\underline{x}=\boldsymbol{G}\underline{a}$ (The shaping domain $B$ will be defined later). Note that using QAM symbols instead of arbitrary integers is equivalent to shifting and scaling the lattice.

We have shown in (\ref{dist_claim}) that the signal code lattice has a better minimum distance than the rectangular lattice of uncoded QAM symbols. However, we still have to show that the density of the signal code lattice points is at least the same as the density of the uncoded symbols lattice. Otherwise, if we have increased the minimal distance at the cost of reducing the lattice density, it is equivalent to scaling the uncoded integers lattice without any coding gain. In order to calculate the density of the lattice points, we shall use the definition of the Voronoi cell of a lattice point, which is defined as the set of all points that are closer to this point than to any other lattice
point. The Voronoi cells of all lattice points are congruent. The volume of the Voronoi cell of a lattice with square generator matrix $\boldsymbol{G}$ is $\det(\boldsymbol{G})$, where for a general $m \times n$ generator matrix $\boldsymbol{G}$ with $m \geq n$ the volume is $\sqrt{\det(\boldsymbol{G}'\boldsymbol{G})}$. %When PAM symbols are used instead of arbitrary integers, each lattice dimension is scaled by 2, but since we compare the lattice density to that of uncoded PAM symbols, this scaling is common to both schemes and thus can be ignored.
Therefore, in order for the signal code lattice to have the same density as the uncoded QAM symbols lattice, we need to scale it by $\left[\det(\boldsymbol{G}'\boldsymbol{G})\right]^{\frac{1}{2N}}$.
Considering the signal code lattice generator matrix (\ref{G_def}), which has a '1' on the main diagonal of its upper $N \times N$ submatrix, and additional $L$ rows below this submatrix, it can be easily seen that for $N >> L$ we have $\left[\det(\boldsymbol{G}'\boldsymbol{G})\right]^{\frac{1}{2N}} \rightarrow 1$. Therefore, no scaling is required. For large $N$, $\boldsymbol{G}$ is a volume preserving transformation, and the signal code lattice points have the same density as a rectangular grid of uncoded QAM symbols. As a result, the improved minimal distance of the signal codes lattice is achieved with the same lattice density as the uncoded symbols lattice, so it has a potential to generate real coding gain.
Though it is well known that trying to achieve good minimum distance is not necessarily the best way to design capacity approaching codes \cite{Forney_Ungerboeck}, we shall use minimum distance as the design criterion, and then test the resulting codes for their probability of error.

%for codin, since the power increase problem can be solved by using appropriate shaping algorithms, as described in Section \ref{shaping}.

We have found an infinite lattice which is good for coding, but encoding by simple convolution results in power increase, as described above. We can solve the power increase problem in the following way: encoding will not be done by direct convolution with the information sequence, but by mapping the information sequence to a lattice point, such that only lattice points that belong to a shaping region will be chosen. This is essentially the shaping region $B$ that was mentioned in the definition of a lattice code above. If the average energy of these selected lattice points is smaller or equal to the average energy of uncoded symbols, then the power increase problem is solved. %This shaping operation is further described in the next section.
Therefore, instead of mapping the information vector $\underline{a}$ to the lattice point $\boldsymbol{G} \underline{a}$, it should be mapped to some other lattice point $\boldsymbol{G} \underline{b}$, such that the lattice points that are used as codewords belong to $B$. The operation of mapping the integer vector $\underline{a}$ to the integer vector $\underline{b}$ is called ``shaping''. 

\begin{figure}
\centering
\vspace {-0.3cm}
\includegraphics[height=2in]{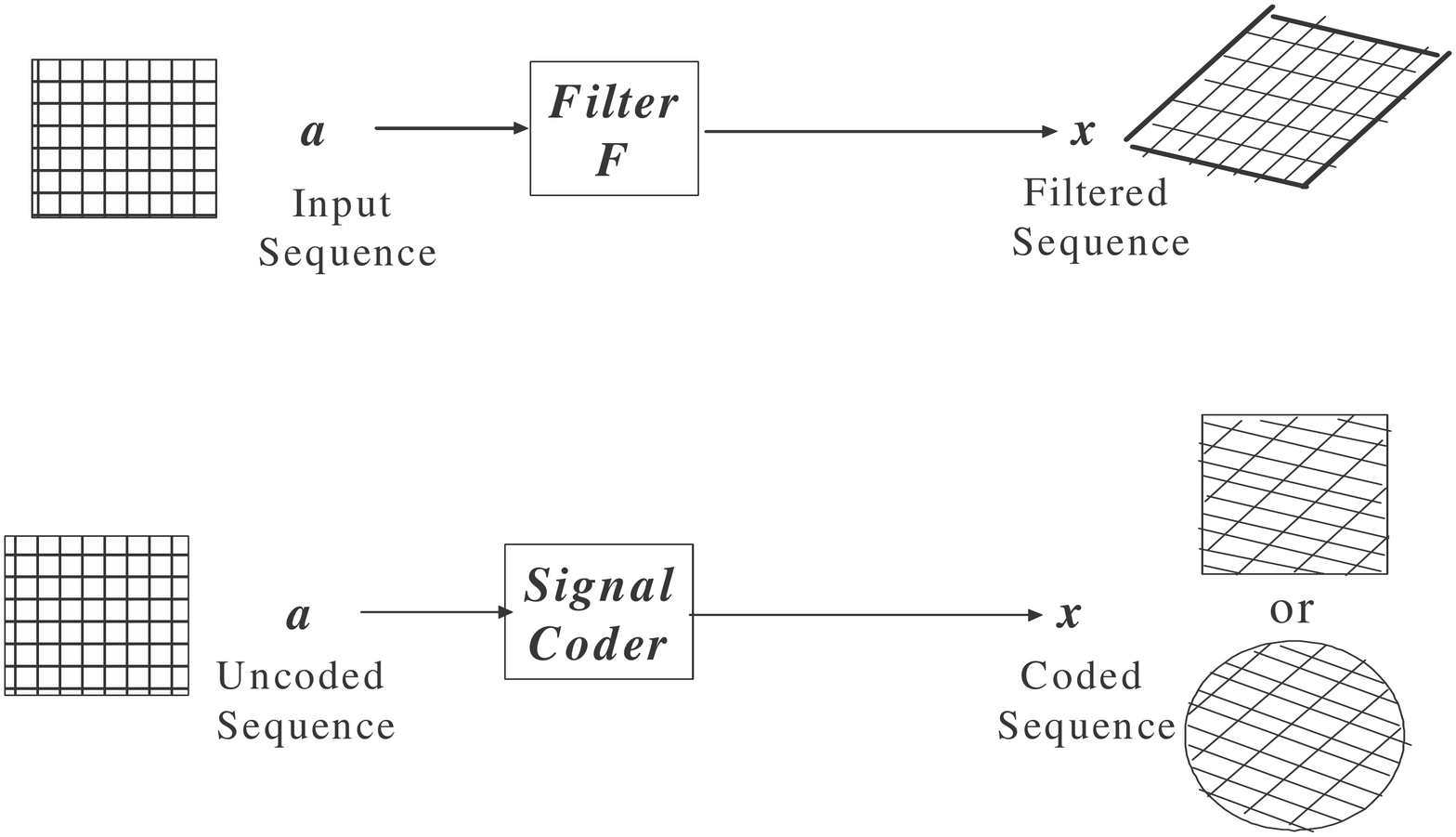}  %, width=2.5in
\vspace {-0.5cm}
\caption{The shaping operation of signal codes}
%\vspace{-0.7cm}
\label{signal_lattice_interp}
\end{figure}

Shaping for signal codes is illustrated in Figure \ref{signal_lattice_interp}. The top part shows how the filtering operation transforms the data sequence from a point on a Cartesian lattice, corresponding to the uncoded signal, to a point on a ``filtered lattice''. On the other hand, the filtering transforms the $(-M,M)^N$ hypercube that contains all the possible $N$-dimensional input vectors into a less power-efficient $(N+L)$-dimensional polytope, and thus increases the signal power. The shaping operation maps the integer information sequence to another sequence such that the output lattice point will be placed inside a shaping region. This shaping region may be, for example, a hypercube or a hypersphere, as shown in the bottom part of Figure \ref{signal_lattice_interp}. In the case of a hypercube, the coded signal will have the same power as the uncoded signal, but with improved packing in the Euclidean space (e.g. larger minimum distance). In the case of a spherical shaping region, in addition to improving the packing, the coded signal's power will be decreased, with a potential shaping gain of 1.53dB relative to the uncoded signal \cite{Forney_Ungerboeck}.

 %the same number of points will be contained in a spherical shaping region for both the signal code lattice and for the uncoded integers rectangular lattice, so both options will carry the same information rate when used as channel codes (except for the factor $\frac{N-L}{N}$, which results from the fact that the uncoded lattice uses all $N$ dimensions, where the signal code uses only $N-L$ dimensions. However, this factor becomes negligible as $N >> L$). We still have to show that for signal codes, the codewords are packed in the Euclidean space such that the probability of error for Gaussian noise is smaller and the resulting coding gain approaches the AWGN channel capacity.

%Shaping can be done by using the nested lattice concept, as described in Section \ref{nested}. However, due to its special convolutional structure, shaping for signal code lattices can be done using simple precoding techniques, as discussed in the next section.

In the next section we shall describe practical shaping algorithms that can be incorporated with the signal code encoding operation.

\section{Shaping} \label{shaping}

\subsection{Tomlinson-Harashima Shaping} \label{Tomli_shaping}
The first shaping method that we shall consider uses a hypercube shaping domain, such that every element $x_n$ of the encoded sequence has real and imaginary parts that belong to the interval $[-M,M)$. %how an encoder can simultaneously perform encoding and shaping. Figure \ref{signal_coder} illustrates a signal code encoder. The input $a_n$ is a sequence of i.i.d. QAM symbols, whose real/imaginary parts are odd integers, uniformly chosen from {-M+1,…,-1,1,…M-1}. 
Assume that the information sequence $a_n$ is a sequence of i.i.d. $M^2$-QAM symbols. The shaping operation maps the symbol sequence into a sequence of extended constellation symbols $b_n$, such that 
\begin{align} 
		b_n = a_n - 2M k_n \label{k_def}
\end{align}
where $k_n$ is a sequence of complex integers. The codeword $x_n$ is then generated by:
\begin{align} \label{x_encode}
		x_n =	 b_n +  \sum_{l=1}^L f_l b_{n-l}
\end{align}
where $F(z) = 1 +  \sum_{l=1}^L f_lz^{-l}$ is the signal code filter pattern of length $L+1$.

Note that as $b_n$ is drawn from the same grid as $a_n$ (the grid of odd integers), the codeword $\boldsymbol{G}\underline{\boldsymbol{b}}$ will also be a lattice point from the same (shifted) lattice as $\boldsymbol{G}\underline{\boldsymbol{a}}$, so this shaping operation preserves the minimal distance and coding gain properties of the lattice $\boldsymbol{G}$. Also, the decoder can recover the information $a_n$ from $b_n$ by a simple modulu $2M$ operation.%it has the same minimum Euclidean distance, $d_b^2 = d_a^2 = 4$. According to Claim \ref{dist_claim}, $d_x^2 \geq d_b^2$, so the combined signal coding and shaping operation increases the minimum distance of the signal. Also, if the decoder recovers the sequence $b_n$, then it can easily reconstruct the original information sequence $a_n$ by a simple modulo operation: $a_n = b_n \mod 2M$.

\begin{figure}
\centering
\vspace {-0.3cm}
\includegraphics[height=2.3in]{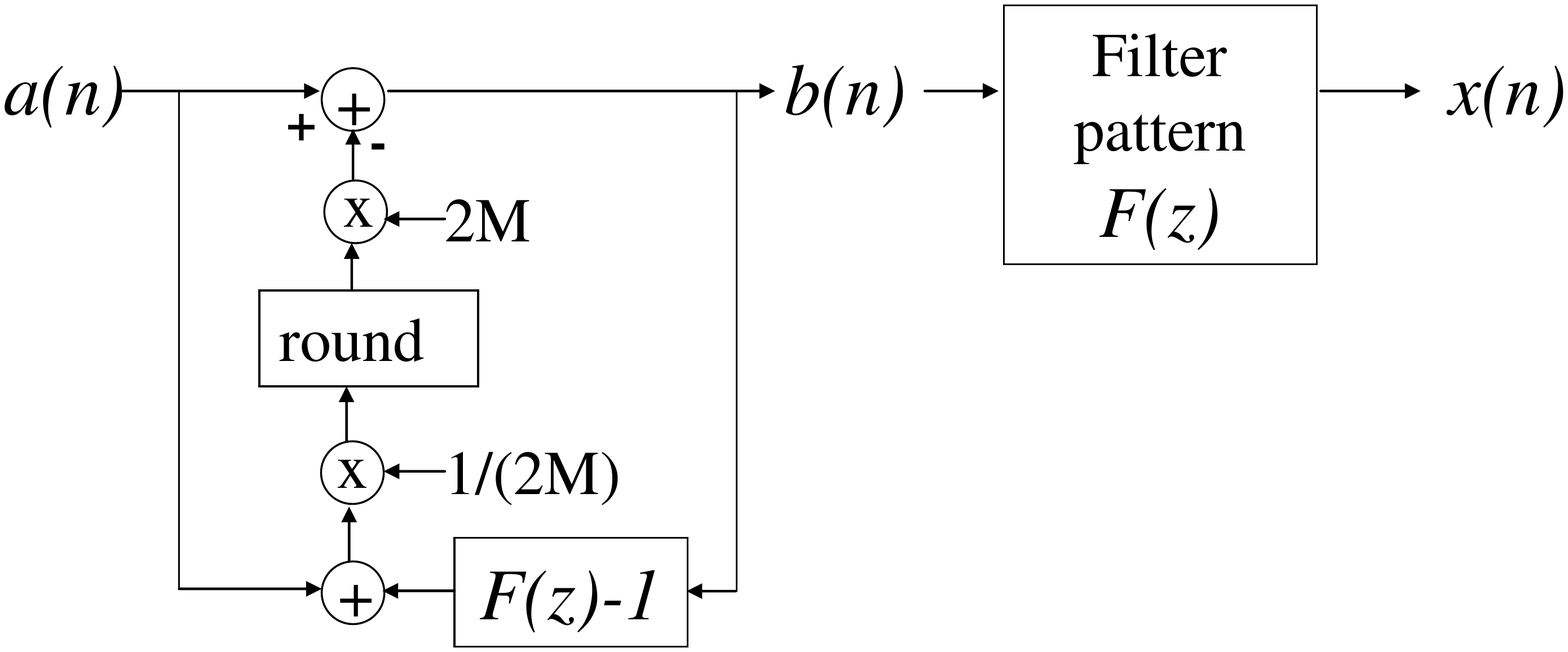}  %, width=2.5in
\vspace {-0.5cm}
\caption{The Tomlinson-based signal encoder}
%\vspace{-0.7cm}
\label{tomli}
\end{figure}

We still have to choose $k_n$ such that the real and imaginary components of $x_n$ are in $[-M,M)$. We have
\begin{align*}
x_n =	 b_n +  \sum_{l=1}^L f_l b_{n-l} = a_n +  \sum_{l=1}^L f_l b_{n-l} - 2M k_n.
\end{align*}
%we can first calculate $a_n +  \sum_{l=1}^L f_l b_{n-l}$. Then, we can add an integer multiple of $2M$ to the real and imaginary parts such that the result will lie in $[-M,M)$. This is 
%equivalent to 
Therefore, the desired result will be achieved by choosing $k_n$ such that:
\begin{align} \label{choose_k_cube}
k_n = \left\lfloor \frac{1}{2M} \left( a_n +  \sum_{l=1}^L f_l b_{n-l} \right) \right\rceil,
\end{align}
where $\left\lfloor x\right\rceil$ denotes the complex integer closest to $x$. The resulting encoder is shown in Figure \ref{tomli}.
%Figure \ref{signal_coder} shows the basic structure of a Signal encoder.

%There are various approaches for choosing $k_n$. The simplest one chooses $k_n$ so that $|x_n|<M$, that is:
Obviously, this approach maps the data into the set of filtered sequences within a hypercube. This mapping has a close relation with precoding and pre-equalization techniques for inter-symbol interference (ISI) channels. In pre-equalization, the transmitter filters the symbols with the inverse channel response, such that after the channel, the signal will have no ISI. However, in order to avoid large transmitted power due to this filtering, precoding is required, and the data symbols are modified as in (\ref{k_def}) such that transmission power will be preserved. We recognize (\ref{k_def})-(\ref{choose_k_cube}) as essentially a Tomlinson-Harashima pre-coder \cite{Tomlinson}, attempting to pre-equalize a phantom linear channel $F^{-1}(z)$.  It is well known \cite{Tomlinson_Forney} that except for some special cases (including for example the case of $F(z)\approx 1$), the output of a Tomlinson-Harashima pre-coder is a spectrally white sequence uniformly distributed over $[-M,M)$, for both real and imaginary parts, so its power is $\frac{2}{3}M^2$. Since the power of uncoded $M^2$-QAM symbols is $\frac{2}{3}(M^2-1)$, the power of $x_n$ is almost the same as the uncoded signal power, albeit higher by a factor of $M^2/(M^2-1)$ which is negligible for large $M$. 

The signal encoding and shaping operations of (\ref{k_def})-(\ref{choose_k_cube}) also resemble commonly used random number generation algorithms. In accordance with Shannon's random coding point of view, signal encoding can be regarded as a transformation of the input data into a pseudo-random sequence.

Note also that the recursive loop of the Tomlinson shaping scheme will be stable (i.e. $b_n$ does not increase without bound) only if the filter $F(z)$ is minimum phase.

\subsection{Flexible Shaping} \label{flexible_shaping}
The Tomlinson-Harashima shaping scheme exploited the equivalence between shaping for signal codes with a hypercube shaping domain, and precoding for a phantom channel $F^{-1}(z)$. In the same manner, any other precoding scheme can be used as well. In this section we shall use flexible precoding \cite{Laroia} for signal code shaping.

For this technique, the shaping operation is:
\begin{align} 
		b_n = a_n - 2 k_n \label{k_def_Laroia}
\end{align}
followed by the standard encoding:
\begin{align} \label{x_encode_Laroja}
		x_n =	 b_n +  \sum_{l=1}^L f_l b_{n-l}.
\end{align}

As for Tomlinson shaping, $b_n$ is drawn from the same grid as $a_n$ (the grid of odd integers), so the codeword $\boldsymbol{G}\underline{\boldsymbol{b}}$ will also be a lattice point from the same (shifted) lattice as $\boldsymbol{G}\underline{\boldsymbol{a}}$. The flexible shaping operation therefore preserves the minimal distance and coding gain properties of the lattice $\boldsymbol{G}$.

The complex integer sequence $k_n$ is now chosen such that the real and imaginary parts of $x_n-a_n$, the difference between the coded and uncoded sequences, will belong to the interval $[-1,1)$. This can be achieved by choosing:
\begin{align} \label{choose_k_cube_Laroja}
k_n = \left\lfloor \frac{1}{2} \left(\sum_{l=1}^L f_l b_{n-l} \right) \right\rceil
\end{align}
With flexible shaping, the coded signal equals the uncoded signal plus an additive ``dither'' signal, whose real and imaginary parts have magnitude less than $1$. Surprisingly, such a small dither signal can yield substantial coding gains, as will be shown in the sequel. Following the same arguments that were used for Tomlinson shaping, this dither would generally be uniformly distributed and uncorrelated with the input sequence. Therefore, the coded signal's real and imaginary parts are uniformly distributed in $[-M,M)$. Also, the information $a_n$ can be recovered from a noiseless codeword by simply slicing (quantizing) the values of $x_n$, so in this sense this coding scheme can be regarded as ``systematic''. In the same manner, the decoder can recover $a_n$ from $b_n$ by generating the codeword elements $x_n$ using (\ref{x_encode_Laroja}), and then slicing $x_n$ to get $a_n$. 

As the coded signal's real and imaginary parts are uniformly distributed in $[-M,M)$, the same $M^2/(M^2-1)$ power increase factor of Tomlinson shaping exists also here. However, unlike Tomlinson shaping, where the coded signal is always mapped to a hypercube, flexible shaping can be combined with standard constellation shaping algorithms, such as trellis shaping \cite{trel_shape} or shell mapping \cite{shell_map}, such that additional shaping gain of 1.53dB can be potentially obtained. This can be done by applying a constellation shaping algorithm to the uncoded sequence $a_n$ prior to flexible shaping and signal encoding. The signal encoding and flexible shaping do not alter the shaping properties of the input signal significantly, since they are equivalent to adding a small dither.

%Note that signal encoding with flexible shaping can be applied to input constellations that reside on an arbitrary non-Cartesian two dimensional lattice in the complex plane (e.g. the hexagonal grid). This is done by replacing the rounding operation in (\ref{choose_k_cube_Laroja}) by an operator that returns the nearest 2-D grid point to a given complex number. In this case, the dither is distributed in the Voronoi region of the 2-D lattice constellation.

\subsection{Nested Lattice Shaping} \label{nested}
The basic Tomlinson and flexible shaping algorithms result in a hypercube shaping domain. As discussed above, it is beneficial to use a spherical shaping domain, since additional 1.53dB of shaping gain can be achieved. However, mapping to a hypersphere is complex, and it is desirable to find simple ways to approximate it.

Consider the Tomlinson shaping operation $b_n = a_n - 2M k_n$. Suppose that instead of setting $k_n$ in a memoryless manner as in (\ref{choose_k_cube}), we choose a sequence $\{k_n\}$ that minimizes the energy of the resulting codeword $\sum_n|x_n|^2$, where $x_n$ is defined in (\ref{x_encode}). Using vector notations, we have
\begin{align} \label{k_def_vec}
\underline{b}=\underline{a}-2M\underline{k}.
\end{align}
Denote the non-shaped lattice point by $\underline{x}'=\boldsymbol{G}\underline{a}$. %, and the shaped lattice point by $\underline{x}=\boldsymbol{G}\underline{b}$. 
From (\ref{k_def_vec}), we then have $\underline{x} = \boldsymbol{G}\underline{b} = \underline{x}' - 2M\boldsymbol{G}\underline{k}$. Choosing $\underline{k}$ that minimizes $\left\|\underline{x}\right\|^2$ is essentially finding the nearest lattice point of the scaled lattice $2M\boldsymbol{G}$ to the non-shaped lattice point $\underline{x}'$, where the chosen codeword $\underline{x}$ is the difference vector between the non-shaped lattice point $\underline{x}'$ and the nearest lattice point $2M\boldsymbol{G}\underline{k}$. Therefore, the codewords will be uniformly distributed along the Voronoi cell of the coarse lattice $2M\boldsymbol{G}$. This is equivalent to a shaping operation with a shaping domain that has the same shape as the Voronoi cell of the lattice, appropriately scaled. It is reasonable to assume that a capacity approaching lattice has a Voronoi cell which resembles a hypersphere, at least from a shaping point of view, so this scheme may attain close-to-optimal shaping gain (compared to uncoded transmission of the original symbols). The resulting shaping scheme is equivalent to nested lattice coding \cite{Zamir_Erez}, where the shaping domain of a lattice code is chosen as the Voronoi region of a different, ``coarse'' lattice, usually chosen as a scaled version of the coding lattice.

\begin{figure}
\centering
\vspace {-0.3cm}
\includegraphics[height=1.7in]{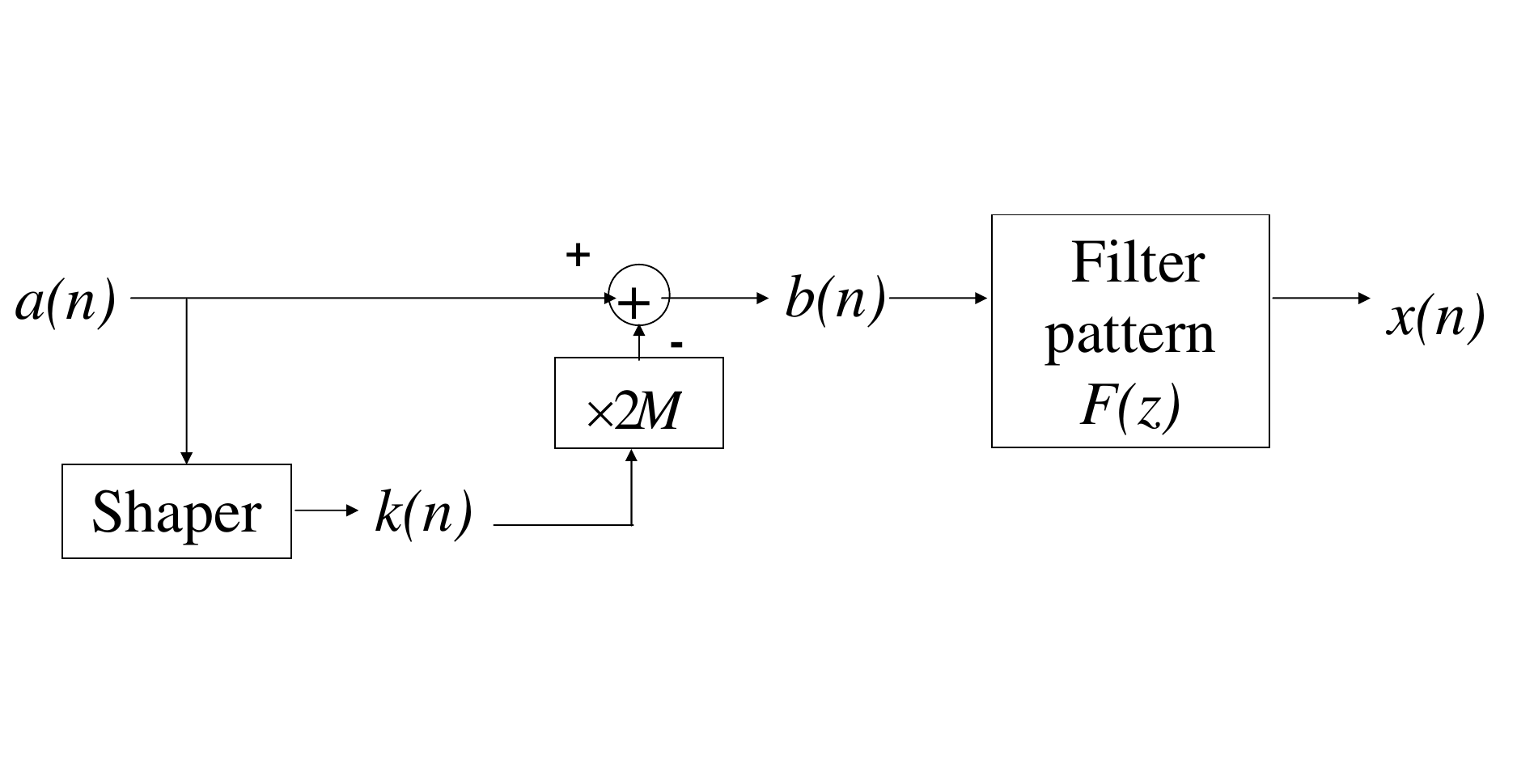}  %, width=2.5in
\vspace {-0.5cm}7
\caption{Nested lattice shaping}
%\vspace{-0.7cm}
\label{signal_coder}
\end{figure}

Finding the closest coarse lattice point $2M \boldsymbol{G} \underline{k}$ to $\underline{x}'$ is equivalent to finding the closest fine lattice point $\boldsymbol{G} \cdot \underline{k}$ to the vector $\underline{x}'/(2M)$.
The complexity of finding the nearest lattice point is the same as the complexity of maximum likelihood decoding in the presence of AWGN. Decoding methods for signal codes will be described in Section \ref{decoders}. However, unlike decoding, for shaping applications it is not critical to find the exact nearest lattice point, as the result will only be a slight penalty in signal power. 
Therefore, approximate algorithms may be considered. As shown in section \ref{sim_res}, close-to-optimal shaping gains can be attained by nested lattice shaping using simple sub-optimal sequential decoders such as the M-algorithm \cite{Aulin_M}.
This algorithm works sequentially on the input symbols of the block, and at each stage stores the M sequences that were found so far with minimum energy. For symbol $n$, each of the M entries is extended with all possible values for $k_n$. Only a finite range of $k_n$ values should be checked, as outside this range the energy of symbol $n$ alone will be large enough such that this path can be truncated immediately. All the extended sequences are sorted, and the M sequences which result in smallest energy are kept as input to the next stage. The value of M determines both the storage and the computational complexity of the shaper. Note that for an M-algorithm with M=1, nested lattice shaping reduces to Tomlinson-Harashima shaping. %Note also that similarly to Tomlinson shaping, the ``inverse shaping'' at the decoder, i.e. transforming from $\underline{b}$ to $\underline{a}$, is a simple modulo calculation: $\underline{a} = \underline{b} \mod 2M$. 
Nested lattice shaping is illustrated in Figure \ref{signal_coder}.

We note that the criterion for choosing $k_n$ can be generalized to meet the needs of communications systems. For instance, the algorithm can combine power optimization with peak magnitude optimization or with short-time power optimization.

\subsection{Terminating the Shaping Operation} \label{termi}
It comes out that all the shaping methods that were presented so far have no natural way to terminate the encoding operation. Even if $a_n$ has finite length and the encoder is fed with zero symbols from a certain point, $b_n$ may continue to be nonzero for a long time. However, the convolution ``tail'' is necessary if we want to maintain the reliability of the last transmitted symbols. Therefore, if we want to partition the data to finite-length blocks, and simply stop the shaping operation abruptly at the end of each block, this convolution tail will have large energy and the resulting codeword will be outside the required shaping domain. A practical solution is to use the shaper and the encoder in a continuous manner (i.e. encode an infinite sequence $a_n$), but at the end of each data block simply transmit the last $L$ values of $b_n$. For the decoder, having this information is equivalent to smooth termination of the encoding. These $b_n$'s should be transmitted such that the probability of error in detecting them should be negligible, compared with regular data transmission. For example, the $b_n$'s can be transmitted using a smaller QAM constellation or a different coding scheme. Transmitting the $b_n$'s results in some overhead, but its impact on code rate becomes negligible as block size increases.

\section{Error Spectrum Analysis for Signal Codes} \label{err_spect}

\subsection{The Error Spectrum and the Union Bound} \label{err_union_bound}

%If we ignore the shaping operation, 
Since signal codes are linear codes, a linear combination of several codewords is itself a codeword. The error performance can then be characterized by the set of all possible error sequences. Note that when the shaping operation is considered, the code is no longer linear. However, the shaping operation chooses a finite subset of the infinite number of possible codewords of a linear code. Therefore, it does not degrade its error performance, so it is sufficient to analyze the error performance of the infinite linear code. 

Each possible error sequence is a convolution of a complex integer error-symbol sequence $\{e_n\}$, whose real and imaginary parts are even integers (including zero), with the filter pattern of the signal code $\{f_n\}$. The Euclidean weight that corresponds to an error-symbol sequence $\{e_n\}$ is defined as the squared Euclidean norm of the resulting error sequence:

\begin{align} \label{err_d_def}
d^2(\underline{\boldsymbol{e}})= \left\|\underline{\boldsymbol{e}} * \underline{\boldsymbol{f}} \right\|^2 = \sum_n |e_n+ \sum_{l=1}^L f_l e_{n-l}|^2
\end{align}

The minimal distance error sequence is the error sequence with at least one nonzero error symbol that has the smallest Euclidean weight. The error spectrum of the code is defined as the sequence of the Euclidean weights of all the possible error symbol sequences.
%Note that the shaping operation chooses a finite subset of the infinite number of possible codewords of the linear signal code. Therefore, it does not degrade the error spectrum of the code or its minimal distance.

Suppose that an optimal Maximum Likelihood (ML) decoder is used for the AWGN channel with complex noise variance $\sigma^2$. Denote by EER (Event Error Rate) the probability that a decoding error starts at a given symbol. The EER can be bounded from above by the union bound \cite{Forney_Ungerboeck}: 
\begin{align} \label{union}
EER  \leq \sum_{\underline{\boldsymbol{e}}} Q\left(\sqrt{\frac{d^2(\underline{\boldsymbol{e}})}{2\sigma^2}} \right)
\end{align}
where the summation in (\ref{union}) is over all possible error-symbol sequences $\underline{\boldsymbol{e}}$ %in the error spectrum 
and Q is the Gaussian error function. %, and $d^2(\underline{\boldsymbol{e}})$ is the Euclidean weight that corresponds to an error-symbol sequence $\underline{\boldsymbol{e}}$. 
Note that for a lattice code, it is enough to sum only over error sequences that correspond to Voronoi-relevant vectors of the lattice, where a Voronoi-relevant vector is a vector that defines a facet of the Voronoi region. See \cite{Agrell_lattice} for methods to check if a lattice point corresponds to a Voronoi-relevant vector or not.

The union bound (\ref{union}) requires summation over a ``practically infinite'' number of error events. At high SNR, the Gaussian Q function decays rapidly as $d^2(\underline{\boldsymbol{e}})$ increases, and thus the union bound can be approximated by taking into account only the low distance error events whose distances fall near the minimum distance. In the sequel, we shall present an algorithm for calculating the low distance error events of signal codes. Its results could therefore be used to evaluate the code performance using the union bound.

Approximating the union bound using the low distance error events does not give a real upper bound or a lower bound to the error probability of the code, but gives only an approximation to an upper bound. Also, it is well known that the union bound may be useless beyond the cutoff rate of the channel \cite{Forney_Ungerboeck}. Therefore, we shall use the approximated union bound as a criterion for choosing the code parameters, but further check is needed to verify the actual code performance.
%Therefore, we can use the low distance error events and the union bound only for codes that do not operate beyond the cutoff rate. However, if the union bound shows that a code can reach the cutoff rate of the channel, it can be regarded as an indication that this code has a potential to be capacity approaching, as the gap between the channel cutoff rate and channel capacity is only 1.7dB for high SNR \cite{Forney_Ungerboeck}. For such codes we shall use extensive simulations in order to check their real performance.
%Simulation results showed that the error rates of the sequential algorithms proposed in the sequel are indeed close to the approximated union bounds. However, these sequential algorithms are non-applicable when the data rate exceeds the channel cutoff rate. 

\subsection{An algorithm for Calculating the Error Spectrum} \label{err_spect_alg_sec}
We shall now present an algorithm that finds all the error sequences whose Euclidean weight is below a given $d_{Search}^2$, where the length of the appropriate error-symbol sequence is smaller than $N_{max}$ symbols.
%The flowchart of the algorithm is shown in Figure \ref{err_spect_flow}. 
The algorithm develops a tree of all possible error sequences, and truncates tree branches as soon as it can identify that all the error events on them will have distances above $d_{Search}^2$. The tree is searched in a Depth First Search (DFS) manner, which can be easily implemented using recursion techniques. The detailed algorithm is presented in Appendix \ref{err_spect_alg_app}.

In fact, This algorithm finds all the lattice points inside a sphere with radius $d_{Search}$. Therefore, it is equivalent to a sphere decoder \cite{Agrell_lattice}, which transforms the lattice generator matrix $\boldsymbol{G}$ to an upper or lower triangular form and then performs a sequential search of the resulting tree. However, the special convolutional structure of the signal code lattice results in an algorithm with reduced computational complexity. Specifically, for signal codes we can eliminate the preprocessing stage of the sphere decoder, which transforms $\boldsymbol{G}$ to a lower or upper triangular form, as the matrix $\boldsymbol{G}$ of (\ref{G_def}) is already in an appropriate form for sequential search. Also, the shift invariance and symmetry properties of the convolution operation, as well as the band-Toeplitz structure of $\boldsymbol{G}$, are used to dilute unnecessary tree branches, as described in Appendix \ref{err_spect_alg_app}.
We also note that the search algorithm is similar to Aulin's algorithm, as presented in \cite{Aulin_err_spect}.

Finding the minimal distance of a lattice code is equivalent to solving the ``nearest lattice point'' problem, which is known to be NP-complete \cite{Agrell_lattice}. However, it comes out that %the special structure of the signal code lattice enables to reduce this complexity significantly, using the branch truncation methods described above, so 
finding the low-distance error events for practical signal code lattices is feasible with the above algorithm. In any case, we can expect the complexity of the algorithm to increase exponentially with $d_{Search}^2$.

\subsection{Filter Patterns with High Coding Gain} \label{patterns}
We shall now use the algorithm of Section \ref{err_spect_alg_sec} to find Filter patterns that generate signal code lattices with large minimum Euclidean distance.
%In this section we present a numerical study of the minimum Euclidean distance of a class of high coding gain Signal codes. 
In a previous work \cite{Anderson_min_dist} it was observed that the best (and worst) linear channels, in terms of optimizing the minimum Euclidean distance under a power constraint, are achieved when all the zeros of the system's Z-transform are on the unit circle of the Z-plane. Motivated by these results, we have performed distance spectrum analysis for filter patterns that have deep spectral nulls, focusing on filter patterns with length $L+1$ that have $L$ zeros at $z=z_0$, i.e. filters of the form:
\begin{align} \label{filter_form}
 F(z) = (1-z_0 z^{-1})^L
\end{align}
where $0<|z_0|<1$. This choice is not necessarily optimal, but the experimental results in the sequel indicate that it can lead to lattices with good coding gain.

In principle, we could have real-valued codewords, using PAM information symbols and filters with real coefficients. However, it turns out that it is better to use QAM symbols and complex valued filter coefficients, as real-valued filters have a drawback that is illustrated by the following example. 

\begin{example}[a 2-tap filter]
Consider a 2-tap filter $F(z) = 1 - f_1 z^{-1}$ with $f_1$ real. In order for $F(z)$ to be minimum-phase we need $|f_1|<1$. It can be easily seen that the minimal distance of the resulting signal code lattice is $4(1+f_1^2)$. Therefore, the asymptotic coding gain is $10\log_{10}(1+f_1^2)$, and it approaches 3dB as $|f_1| \rightarrow 1$. Consider the case $f_1=0.99$, i.e. $F(z) = 1 -0.99 z^{-1}$. This is a high-pass filter with a notch in its frequency response, centered at 0Hz. Assume that a long sequence of symbols with constant value is filtered with this filter. As a constant-valued sequence has most of its energy located at 0Hz, it will be strongly attenuated by the notch of the filter, resulting in a low-energy output. Therefore, extending a given error-symbol sequence by duplicating its last symbol several times will generate an error event with only slightly higher weight than the one that corresponds to the original sequence. As a result, the error spectrum of this code will contain a large number of low-distance error events, with weight which is close to the minimal distance. It comes out that the improvement of $d_{min}$ is 2.97dB, but there are more than 20,000 error events within 1dB of the minimal distance error event. This is certainly undesirable, and resembles the phenomenon of catastrophic error events in binary convolutional codes. 

This problem can be avoided by choosing complex-valued $f_1$ with a nonzero complex phase, e.g. $\pi/4$, and an amplitude that is a bit smaller than 1. This choice yields the expected 3 dB gain, but this time without the above singularity. 
\end{example}

%For this purpose, we employed Aulin's distance spectrum analysis algorithm \cite{Aulin_err_spect} that finds all the error sequences whose squared Euclidean norm is below a given $d_{SEARCH}^2$ and whose length is smaller than $N_{SEARCH}$ symbols. This algorithm develops a tree of all possible error sequences, and prunes tree branches as soon as it can identify that all the error events on them will have distances above $d_{SEARCH}^2$. In effect, this algorithm is equivalent to a sphere decoder \cite{Agrell_lattice} that finds all the lattice points that lie within a sphere of radius $d_{SEARCH}$ around the origin. However, it uses the special convolutional structure of the Signal codewords in order to reduce the computational complexity.
As a result, we shall focus on signal codes with complex coefficients.
The error spectrum calculation algorithm was then used to find the minimum Euclidean distance for various values of the magnitude and phase of the complex zero $z_0$ of (\ref{filter_form}), and the filters with the largest minimum distance gain that were found are shown in Table \ref{dmin_table}. For each filter, the second column of the table shows the minimum squared Euclidean distance, and the third column shows $N_{min}$, the length of the minimum distance error event in symbols. The last column shows the amount of increase in $d_{min}^2$, relative to an uncoded QAM signal, measured in dB (recall that the minimum distance for an uncoded QAM signal is 4).
It can be seen that even a short filter with only $L=2$ zeros can achieve considerable improvement of over 5 dB in $d_{min}^2$. Also, the improvement in $d_{min}^2$ grows as the spectral null of the encoder's filter becomes deeper, either by increasing the number of zeros $L$ or by letting the zero $z_0$ approach the unit circle more closely. %Note, however, that increased $d_{min}^2$ does not guarantee increased coding gain, though it can certainly be an indication to it, as already discussed in Section \ref{}.

The gap between uncoded transmission and channel capacity is about 9dB for bit error rate of $10^{-6}$, where 1.5dB correspond to shaping gain \cite{Forney_Ungerboeck}. The gap between channel capacity and channel cutoff rate is approximately 1.7dB for high SNR. Therefore, the required coding gain for reaching the channel cutoff rate is 9-1.5-1.7 = 5.8dB. As the $d_{min}^2$ improvement of all the codes in Table \ref{dmin_table} is 5.7dB and above, the approximated union bound of Section \ref{err_union_bound} indicates that these codes can reach the channel cutoff rate and have a potential to work beyond it. Indeed, the simulation results of Section \ref{sim_res} demonstrate operation of the code that corresponds to the fourth row of Table \ref{dmin_table} at approximately 1dB from channel capacity.

\begin{table}
\caption{High Coding Gain Filter Patterns}
%\vspace{-0.3cm}
\label{dmin_table}
\centering
%\begin{tabular}{p{4.5cm}||p{2.4cm}}
%\hline \hline
%Conditions on Node & Action to be Taken \\
%\hline \hline
%\end{tabular}
\begin{tabular}{p{3.0cm}|p{1.3cm}|p{1.2cm}|p{1.5cm}}
 $F(z)$ &  $d_{min}^2$ &  $N_{min}$ [symbols] &  $d_{min}^2$    improvement [dB] \\
\hline \hline
 $(1+0.90e^{j\pi/8}z^{-1})^2$ &  14.81 &  3 &  5.7dB \\
\hline
 $(1+0.98e^{j\pi/8}z^{-1})^2$ &  17.33 &  3 &  6.4dB \\
\hline
 $(1+0.95e^{j\pi/8}z^{-1})^3$ &  20.53 &  10 &  7.1dB \\
\hline
 $(1+0.98e^{j0.09\pi}z^{-1})^3$ &  23.59 &  5 &  7.7dB  \\
\hline
 $(1+0.95e^{j0.08\pi}z^{-1})^4$ &  31.27 &  12 &  8.9dB \\
\hline
\end{tabular}
%\vspace {-0.7cm}
\end{table}

It is interesting to note that for all the patterns of Table \ref{dmin_table}, the minimal-distance error-symbol sequence $\underline{\boldsymbol{e}}_{min}$, whose length is denoted by $N_{min}$, satisfied $e_{min}(t) = e_{min}(N_{min}-t)$, for $1 \leq t \leq N_{min}/2$, up to a possible complex conjugate operation and rotation by 0, 90, 180, or 270 degrees (where the conjugate and rotation operations are independent of $t$).

\subsection{Properties of the Error Spectrum}

The algorithm of Figure \ref{err_spect_flow} was run for the signal code lattice based on the fourth filter pattern of Table \ref{dmin_table}, with $N_{max}=1000$ and $d^2_{Search}=42$. As explained above, the complexity of the distance spectrum algorithm increases exponentially with $d_{Search}^2$. Indeed, in order to generate the error spectrum for the above conditions, 500 billion tree nodes had to be examined.

%Figure \ref{err_spect_low_events} shows the low-distance error events of the error spectrum. The leftmost line corresponds to the minimum-distance error event. It can be seen that the error specturm behaves regularly, and there is no ``flood'' of error events above the minimal distance.

\begin{figure}
\centering
%\vspace {-0.3cm}
\includegraphics[width=3.5in]{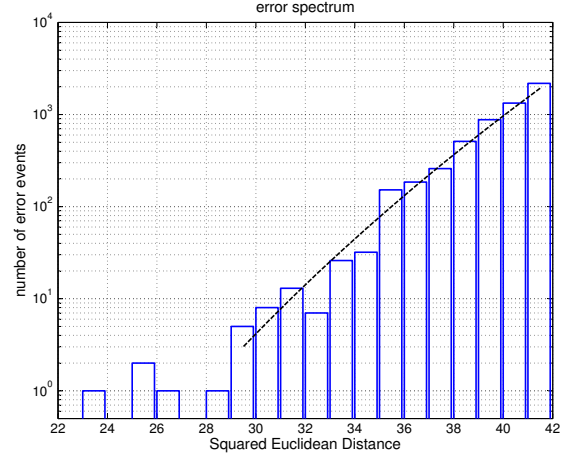}  %, width=2.5in
%\vspace {-0.5cm}
\caption{Error spectrum histogram for a signal code lattice. The dashed line shows a least square fit to an exponential model.}
%\vspace{-0.7cm}
\label{err_spect_hist}
\end{figure}

Figure \ref{err_spect_hist} shows a histogram of the squared Euclidean distance of the error events, where each bar corresponds to an interval of length $1$ and shows how many error events had weight within this interval. Note that shifts and complex rotations by an integer multiple of 90 degrees of error events are not counted separately. 

The leftmost bar corresponds to the minimum-distance error event whose weight is $23.59$. It can be seen that the first several low-distance error events (weights $23$-$29$) are discrete events, and there is no ``flood'' of error events above the minimal distance. For higher weights ($30$ and above), the number of error events grows exponentially with the Euclidean weight, as seem from the least square fit of the histogram to the exponential model $N(d)=\alpha d^\beta$, which is plotted in the figure with a dashed line.

%\begin{figure}
%\centering
%%\vspace {-0.3cm}
%\includegraphics[width=3.5in]{err_spect_low_events.pdf}  %, width=2.5in
%%\vspace {-0.5cm}
%\caption{Low-distance error spectrum for a signal code lattice.}
%%\vspace{-0.7cm}
%\label{err_spect_low_events}
%\end{figure}

From a first glance, the exponential behavior is surprising. We expect the number of error events with weight lower than $d^2_{Search}$ to be proportional to the total number of lattice points within a hypersphere of radius $d_{Search}$. This number is approximately proportional to the volume of the hypersphere, which is polynomial in $d_{Search}$. Therefore, we should expect polynomial behavior from the error spectrum, and not exponential behavior. However, we have used a lattice dimension which is much higher than the search radius. As a result, all error events have most of their error symbols equal zero (note that the minimum weight of an error event whose symbols are all nonzeros is $4N_{max}=4000$ which is much larger than $d_{Search} = \sqrt{42}$). For this situation, when the search radius is further increased, new events can be generated by using sequences with more nonzero symbols (e.g. by extending existing sequences with another symbol), and this explains why their number grows exponentially. For large search radius, however, all the symbols of the error sequences are nonzero, so as the distance increases, it is not possible to add nonzero symbols, but the existing symbols have to be increased, resulting in polynomial behavior.

See Appendix \ref{Cartesian_calc} for an analytic calculation of the error spectrum for the simple Cartesian lattice with infinite dimension, which demonstrates the exponential nature of the error spectrum which was described above.
%we have calculated the error spectrum with search radius $d^2_{Search}=42$ using a dimension of $N_{max}=1000$. Note that $4N_{max}$ is the largest Euclidean weight of all points that are located within a hypercube of dimension $N_{max}$ with length 2 for each dimension. This is also the minimum weight of an error event whose symbols are all nonzeros. As $d^2_{Search} << 4N_{max}$, we have practically calculated the error spectrum of an infinite lattice. Therefore, the exponential behavior is observed for error events whose distance is low, relative to the lattice dimension. However, For the error spectrum of a lattice with fixed finite dimension and asymptotically large $d^2_{Search}$, 

%The different behavior for low and high distance events can be qualitatively explained as follows. For low distance events, the number of nonzero symbols that generate the events is much smaller than the lattice dimension. Therefore, when the distance is increased, new events can be generated by using sequences with more nonzero symbols, so their number grows exponentially. For high distance, all the symbols of the error sequences are nonzero, so as the distance increases, it is not possible to add nonzero symbols, but the existing symbols have to be increased, resulting in polynomial behavior.

\subsection{The Narrow band Nature of the Shaped Symbols $b_n$} \label{NB_nature}

For signal codes, the codeword $x_n$ is generated by convolving the shaped symbols $b_n$ with the filter pattern $f_n$. Therefore, the shaped symbols $b_n$ can be regarded as the output of the filter $1/F(z)$ whose input is the transmitted sequence $x_n$. It was shown in Section \ref{patterns} that good filter patterns have deep spectral nulls. Therefore, the filter $1/F(z)$ is a narrow band bandpass filter, whose passband frequencies depend on the phase of the zeros of $F(z)$. For Tomlinson shaping and flexible shaping, $x_n$ is approximately an i.i.d sequence with a flat power spectrum. Therefore, $b_n$ will be a narrow band signal, whose energy depends on the gain of the narrow band filter $1/F(z)$, which in turn depends on the depth of the notch of the filter pattern $F(z)$.

If the gain of $1/F(z)$ is large, the dynamic range of the $b_n$'s will be also large, and many bits will be needed to store them. When data is transmitted in blocks with finite length, the $L$ last $b_n$'s of each block should be transmitted at the end of the block (See Section \ref{termi}). Therefore, it is desirable that the $b_n$'s will be stored in less bits. The narrow band nature of the sequence $b_n$ can be used to achieve it, as described in the sequel.

As an example, consider the filter pattern of the fourth line of Table \ref{dmin_table}: $F(z)=(1+0.98e^{j0.09\pi}z^{-1})^3$. For this filter pattern, it comes out that when the information symbols $a_n$ belong to a 64-QAM constellation, and Tomlinson shaping is used, the real and imaginary parts of the $b_n$'s have a dynamic range of 17 bits (each). Therefore, $17 \times 2 \times 3 = 102$ bits are required to store the last 3 $b_n$'s of each block. However, as the $b_n$'s are the output of a narrow band filter, they can be easily ``compressed''. Instead of transmitting $b_1,b_2,b_3$ we can transmit $b'_1, b'_2, b'_3$ where $b'_1=b_1$,  $b'_2$ is the prediction error of predicting $b_2$ from $b_1$ using the prediction error filter $(1+0.98e^{j0.09\pi}z^{-1})$, i.e. $b'_2=b_2+0.98e^{j0.09\pi}b_1$, and $b'_3$ is the prediction error of predicting $b_3$ from $b_1$, $b_2$ using the prediction error filter $(1+0.98e^{j0.09\pi}z^{-1})^2$. The dynamic range of $b'_1$ is still 17 bits, but the dynamic range of $b'_2$ and $b'_3$ is now 12 and 7 bits, respectively. Therefore, the total number of bits required to store 3 consecutive $b_n$'s is now only $(17+12+7) \times 2 = 72$ bits.

%(1+0.90ej?/8z-1)2	14.81 (3)	5.7
%(1+0.98ej?/8z-1)2	17.33 (3)	6.4
%(1+0.95ej?/8z-1)3	20.53 (10)	7.1
%(1+0.98e0.09j?z-1)3	23.59 (5)	7.7
%(1+0.95e0.08j?z-1)4	31.27 (12)	8.9
%
%

\section{Computationally Efficient Decoders} \label{decoders}

\subsection{Reduced Complexity Maximum-Likelihood Decoding} \label{reduced_comp}
%Straight-forward optimal maximum likelihood decoding for signal codes is computationally complex. In this section we develop approximated maximum likelihood decoders, where simulation results for their performance is provided in Section \ref{sim_res}. 

Let the transmitted codeword be $x_n$ of (\ref{x_encode}), and consider the additive white Gaussian noise (AWGN) channel $y_n=x_n+w_n$, where $w_n$ is a sequence of zero-mean, i.i.d complex Gaussian random variables with variance $\sigma^2$. The optimal ML decoder should maximize
\begin{align} \label{ml_max}
	L(\underline{y}|\underline{a}) = -\sum_n\left|y_n - \sum_{l=0}^L f_l b^{\boldsymbol{a}}_{n-l}\right|^2
\end{align}
where $b^{\boldsymbol{a}}_n$ is the sequence of shaped symbols that correspond to $\underline{a}$. 

However, it is not simple to take the non-linear shaping operation into account in the decoding process. We therefore propose to use ``lattice decoding'' \cite{Zamir_Erez}. In lattice decoding, the decoder ignores the shaping operation and decodes to the infinite lattice, i.e. it finds the nearest lattice point to the received noisy codeword. With proper coding and decoding schemes, channel capacity can still be approached although lattice decoding is used \cite{Zamir_Erez}. 

Therefore, we shall refer to the $b_n$'s as free variables and look for the ``Quasi Maximum Likelihood'' (QML) sequence $\underline{b}_{QML}$ that maximizes (\ref{ml_max}) over all values of the $b_n$'s:

\begin{align} \label{ml_max_b}
\underline{b}_{QML} = \arg \max_{\underline{b}}	L(\underline{y}|\underline{b})
\end{align}
where
\begin{align} \label{free_b_max}
L(\underline{y}|\underline{b}) =  -\sum_n \left|y_n - \sum_{l=0}^L f_l b_{n-l}\right|^2
\end{align}

The data sequence $a_n$ is then estimated by performing the inverse shaping operation on the detected $\underline{b}_{QML}$. For the Tomlinson and nested lattice shaping algorithms, the inverse shaping is simply taking the modulo $2M$ value of $\underline{b}_{QML}$. For flexible shaping, the inverse shaping operation first re-generates the codeword $\underline{x}$ from $\underline{b}_{QML}$, and then quantizes to the nearest QAM symbol, as described in Section \ref{shaping}. 

When we apply lattice decoding to signal codes, we essentially face an equalization problem: a QAM symbol sequence $b_n$ was convolved with a filter pattern, and has to be detected from the noisy convolution output. %We propose a simpler, quasi-maximum likelihood sequence decoding (QMLSD) approach. 
As shown in \cite{Forney_MLSE}, this can be implemented by a Viterbi Algorithm (VA) whose state is $(b_{n-1},..., b_{n-L})^T$. 
The number of trellis branches of the proposed VA is equal to the constellation size of $b_n$, raised to the power of $L$. Therefore, the VA is practical only if $L$ and $M$ are small, and if the constellation expansion is not prohibitively high. The constellation size of $b_n$ is at least $M^2$ (the constellation size of $a_n$), but it may be much larger, depending on $F(z)$ and on the shaping algorithm. It comes out that good codes generate large $b_n$ values. In general, a straightforward VA may be too complex, and it is beneficial to find a reduced-complexity VA decoder.

Reduced complexity Viterbi decoding can follow the well-known techniques used in the context of convolutional codes and maximum likelihood channel equalization. One class of such techniques is sequential decoding, e.g., the Fano \cite{Gallager_Fano} and stack \cite{Viterbi_Omura} algorithms. Another class includes list algorithms such as the M-algorithm \cite{Aulin_M} and the T-algorithm \cite{Anderson_T}. A third class is Reduced States Sequence Detection (RSSD) algorithms (e.g. \cite{RSSE}). 

All these methods try to reduce complexity by searching only a part of the full tree which is spanned by the Viterbi decoder.
As a result, these algorithms suffer from Correct Path Loss (CPL) events, in which the true trellis path is excluded from the ``short list'' of paths that the algorithm maintains. These events are characterized by long (sometimes very long) error bursts. However, if the data is partitioned to finite-length blocks such that decoding can start again for every block, and if the main performance measure is frame error rate and not bit error rate (i.e. if a block has errors, it does not matter how many), then this effect is not a problem.
%Another advantage of partitioning the data to finite length blocks is the possibility to further reduce the computational complexity by using bidirectional decoding, as described below.

However, the reduced complexity decoding algorithms have a more severe problem. In a classical paper \cite{Berlekamp_Jacobs}, Jacobs and Berlekamp have shown that the computational complexity of sequential decoding of any tree code obeys a Pareto distribution. Such a distribution results in the computational cutoff effect, where for a given information rate, complexity increases abruptly below some cutoff SNR, where the variance and/or the mean of the number of computations becomes infinite. Therefore, all the above reduced-complexity decoders are expected to be effective only above the cutoff SNR, which is known to be approximately 1.7dB above the Shannon capacity for the high SNR regime of the AWGN channel \cite{Forney_Ungerboeck}.

On the other hand, even when the mean or the variance of the number of computations becomes infinite, the probability that this number will exceed a pre-defined threshold is still finite. Therefore, if a target finite error rate is defined, sequential decoders can achieve this error rate with finite (and probably large) complexity even beyond the cutoff rate. %This complexity can be reduced by using bidirectional sequeintial decoding, as described in the next subsection.
In Section \ref{sim_res} we shall show that the sequential stack decoder can be used for simple and effective decoding of signal codes close to the cutoff rate. We shall also use bidirectional sequential decoders with large complexity to demonstrate that small finite error rate can be achieved even 0.5dB beyond the cutoff rate, with large (but still finite) computational resources. %This gives a strong indication that the signal code lattice is indeed a capacity approaching lattice. %Finding a practical decoder for signal codes beyond the cutoff rate is certainly a topic for further research.

We shall now turn to describe the stack decoder and its application to signal codes, and then show how it can be used in a bidirectional decoding scheme.

\subsection{The Heap-based Stack Decoder} \label{stack_dec}
The stack decoder \cite{Viterbi_Omura} is a simple and effective algorithm to decode tree codes. A stack of previously explored paths is initialized with the root of the tree code. At each step, the path with best score in the stack is extended to all its successors, and then deleted from the stack. The successors then enter the stack. For a finite block with known termination state, the algorithm terminates when a path in the stack reaches the termination state at the end of the block.

In principle, an infinite stack is required, as the number of paths continuously increases. Practically, a finite stack must be used, so whenever the stack is full, the path with worst score is thrown away.
Therefore, a practical stack decoder should find at each step the paths with best score and worst score in the stack.
%The stack algorithm is required to sort all the paths in its memory and find the one with the best score in order to extended it. In case the memory is full, the algorithm also needs to find the path with the worst score in order to remove it. %The sorting operation can generate a heavy computational burden.

\begin{figure}
\centering
%\vspace {-0.3cm}
\includegraphics[width=2.5in]{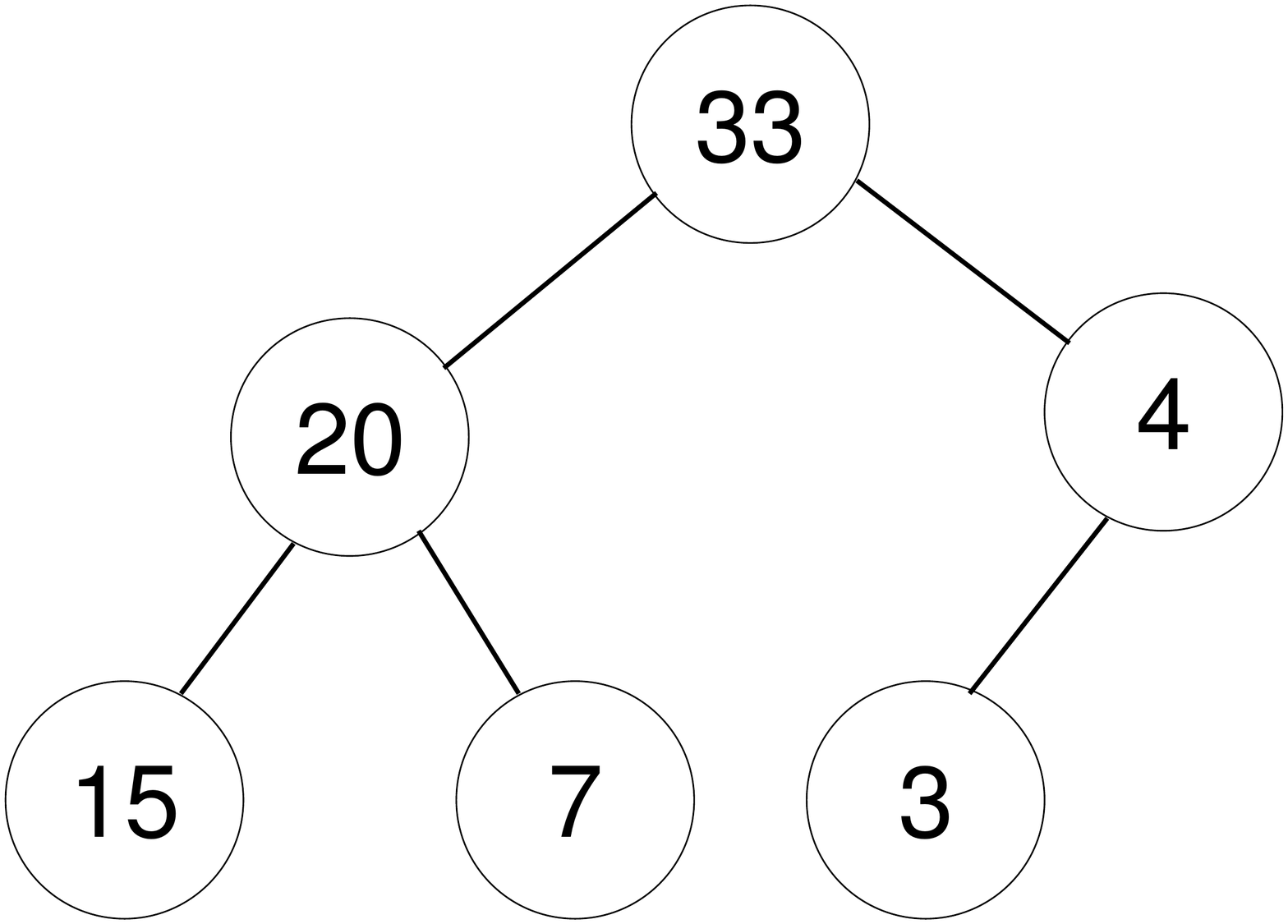}  %, width=2.5in
%\vspace {-0.5cm}
\caption{An example of the heap data structure.}
%\vspace{-0.7cm}
\label{heap_example}
\end{figure}

We propose an efficient implementation of the stack algorithm using the heap data structure \cite{Cormen}. This implementation is suitable for any use of the stack decoder, not necessarily for signal codes.
%As well as we know, such an implementation of the stack algorithm was not proposed before.
A heap is a data structure that stores the data in a semi-sorted manner (See an example in Figure \ref{heap_example}). Specifically, data is arranged in a binary complete tree (i.e. all the levels of the tree are populated, except for the lowest level, whose populated elements are located consecutively at the leftmost locations). The value of each node is larger or equal to the value of its successors. Practically, the heap is stored in a linearly-addressed array, without any overhead (i.e. the root of the tree is stored in location 0 of the array, the two elements of the second level are stored in locations 1 and 2, the four elements of the third level at locations 3,4,5,6 and so on). The parent node of the element at location $i$ of the array is stored at location $\left\lfloor \frac{i-1}{2}\right\rfloor$, and its two children are at locations $2i+1$ and $2i+2$, where $\left\lfloor x\right\rfloor$ denotes the largest integer smaller than $x$.

In order to insert a new element to the stack, the element is initially inserted at the lowest level of the tree, adjacent to the rightmost current element. Then, the new element is moved up the path toward the root, by successively exchanging its value with the value in the node above. The operation continues until the value reaches a position where it is less than or equal to its parent, or, failing that, until it reaches the root node. 

Extracting the maximum element is simple, as the maximum is always at the root of the heap. However, in order to maintain a complete tree, the following procedure is used to delete the maximal element from the stack. First, the root element is deleted and replaced by the rightmost element of the bottom level of the tree. Then, its value is moved down the tree by successively exchanging it with the larger of its two children. The operation continues until the value reaches a position where it is larger than or equal to both its children, or, failing that, until it reaches a leaf. 

It can be easily seen that for a stack of size $n$, extracting the minimum or inserting a new element requires $O(\log_2n)$ operations. As noted above, a practical implementation of the stack algorithm requires to efficiently extract both the minimal and the maximal elements at each step. The deap \cite{deap} or min-max heap \cite{min_max_heap} are modified versions that allow to extract either the maximum or the minimum with $O(\log_2n)$ operations. These data structures are therefore suitable to hold the stack; otherwise, at least $O(n)$ operations may be required to extract the minimum or the maximum, which may dominate the computational load of the algorithm.

Note that for the Tomlinson-Harashima and flexible shaping methods, we can reduce the computational complexity of the stack algorithm by incorporating shaping information to the decoding (In this sense, it is no longer lattice decoding, as defined in Section \ref{reduced_comp}). Specifically, for these shaping algorithms we know that the codeword elements are bounded, since $|x_n| < M$. Therefore, for every path of length $n$ in the stack, we can calculate the resulting symbol $x_n$, and if $|x_n| > M$ we can immediately truncate this path. This technique is very effective for complexity reduction, and will be referred to as ``x-range testing''.

For decoding of signal codes, each entry in the stack should include a score (by which the heap is organized) and a list of $b_n$ symbols that define the path in the code tree. As the codeword may be long (e.g. 1000 symbols), storing the path elements requires a large amount of memory. However, this amount can be reduced as follows. In general, a path in the stack starts in the root of the code tree. Then, it follows the correct path for several symbols, and diverges from it at a certain point. As a path diverges from the correct path, it begins to accumulate score at a much higher average rate than the correct path. Therefore, paths that diverged from the correct path for many symbols will have much worse score than the correct path, and will be thrown away from the stack with high probability. As a result, most of the paths in the stack will have a common start, which equals the first symbols of the correct path, and will differ only at the last few symbols. This observation also holds for Viterbi decoding of convolutional or trellis codes, where in principle, a decision for a data symbol can be taken only after the decoder reached the end of the frame. Practically, decisions are taken by back-tracking the best path for a finite number of symbols to the past, where all the paths are assumed to converge. The same can be done here, where each entry in the stack will only hold the several last symbols of the path, and decision is taken for the older symbols. The stored length should be chosen such that the additional error probability due to these early decisions will be negligible.

However, this method is still not optimal, as most of the paths diverged from the correct path for a small number of symbols, but equal storage is allocated for all paths according to the worst case paths that might have diverged for a larger period. This can be improved as follows. Instead of storing a separate path for each stack entry, all the paths are stored together in a ``symbol memory'', using linked lists of data symbols. Each entry of the symbol memory stores a data symbol $b_n$ and a link to another entry. It also stores the number of entries that are linked to this entry. Each score entry in the stack is linked to the last (newest) symbol of the corresponding path, which is stored in the symbol memory. This symbol is linked to the previous symbol in the path, and so on. The path of each stack entry can be simply followed by back-tracking the links until the root. In order to maintain this database, whenever a path enters the stack after deletion of its parent, a new data symbol is added to the symbol memory, storing the last data symbol of the new path, and a link to the last symbol of the path of its parent entry (which is not deleted from the symbol memory when the parent node is deleted from the stack). A symbol is deleted from the symbol memory only when no other symbol is linked to it. This way, the minimal number of symbols is stored at each point, and memory usage is optimized. Similarly to the previous approach, storage should be allocated to the symbol memory such that the additional error probability due to symbol memory overflow is negligible.

We have still not addressed the problem of assigning scores to the paths in the stack. Naturally, we would assign scores to the paths in the stack according to their likelihood (\ref{free_b_max}). However, the stack contains paths of different lengths. If we use (\ref{free_b_max}), shorter paths will get higher score, as less negative terms are accumulated. This is not desired, since we want to extend the path which coincides with the correct path, even if it is much longer than other incorrect paths in the stack. Therefore, the path scores should be defined such that the effect of path length is eliminated. This problem is addressed in the next subsection.

\subsection{The Fano Metric} \label{Fano_metric_sec}

For sequential decoding of binary convolutional codes, Fano suggested to subtract a bias term from each increment of the natural likelihood score, where the bias equals the code rate $R$. Massey \cite{Fano_metric} has shown that the score assignment problem is equivalent to decoding of a code with variable length codewords, and that the Fano metric is indeed the correct choice for stack and Fano decoding of binary convolutional codes, in the sense that the most likely path is extended in each step.

Massey's derivation can be extended to the Euclidean case%in a straightforward manner
, as done in \cite{TVZ_preprint} for the general case of lattice decoding. Here, we follow the lines of \cite{TVZ_preprint} and develop the Fano metric for signal codes with Tomlinson-Harashima shaping. It comes out that similarly to convolutional codes, in order to extend the most likely path in each step, a bias term has to be subtracted from the score increments of (\ref{free_b_max}):

\begin{align}\label{ml_max_Fano}
L(\underline{y}|\underline{b}) =  -\sum_n \left[\left|y_n - \sum_{l=0}^L f_l b_{n-l}\right|^2 - B\right]
\end{align}

%\begin{align} \label{metric_body}
%\boldsymbol{L}(\boldsymbol{x}_m, \boldsymbol{y}) = \sum_{i=0}^{n_m-1}\left[-(y_i-x_{m,i})^2 + B\right]
%\end{align}
where:
%\begin{align} 
%B \buildrel \Delta \over = \sigma^2 \cdot \log \frac{2}{\pi \sigma^2}
%\end{align}

\begin{align} \label{B_def_body}
B \approx \sigma^2 \cdot \log \frac{4}{\pi \sigma^2}
\end{align}
See Appendix \ref{Fano_app} for the derivation of (\ref{ml_max_Fano}) and (\ref{B_def_body}).

We can make an interesting observation from (\ref{B_def_body}). In order for the stack algorithm (as well as the Fano algorithm) to work, the expected value of the correct path must increase along the search tree, where it must decrease for the incorrect paths \cite{Gallager_Fano}. For the correct path, we have $E\{\left|y_n - \sum_{l=0}^L f_l b_{n-l}\right|^2\} = \sigma^2$. Therefore, in order for the expected value of the path score to increase along the tree, we need to have $B>\sigma^2$ in (\ref{ml_max_Fano}). From (\ref{B_def_body}), we then have $\log\left(\frac{4}{\pi\sigma^2}\right)>1$, resulting in $\sigma^2<\frac{4}{\pi e}$. 

Now, when using a lattice code for the real-valued AWGN channel with power limit
$P$ and noise variance $\sigma ^2$, the maximal information rate
is limited by the capacity
$\frac{1}{2}\log_2(1+\frac{P}{\sigma^2})$. Poltyrev
\cite{Poltyrev} considered
%the case of infinite constellations for
the AWGN channel without restrictions.  If there is no power
restriction, code rate is a meaningless measure, since it can be
increased without limit. Instead, it was suggested in
\cite{Poltyrev} to use the measure of constellation density,
leading to a generalized definition of the capacity as the maximal
possible codeword density that can be recovered reliably. When
applied to lattices, the generalized capacity implies that there
exists a lattice $\boldsymbol{G}$ of high enough dimension $n$
that enables transmission with arbitrary small error
probability, if and only if $\sigma^2 < \frac{\sqrt[n]{|det(\boldsymbol{G})|^2}}{2 \pi e}$. %For
%the high SNR regime,
A lattice that achieves the generalized
capacity of the AWGN channel without restrictions, also achieves
the channel capacity of the power constrained AWGN channel, with a
properly chosen spherical shaping region (see also \cite{Zamir_Erez}).

%Now, it was shown by Poltyrev \cite{Poltyrev} that there
%exists a real lattice with a square non-singular generator matrix $\boldsymbol{G}$ of high enough dimension $n$
%that enables transmission with arbitrary small error
%probability for the AWGN channel without power constraints, if and only if the Gaussian noise variance satisfies $\sigma^2 < \frac{\sqrt[n]{|det(\boldsymbol{G})|^2}}{2 \pi e}$. It was also shown that 
As the signal code lattice is a volume preserving transformation of the rectangular lattice, and our basic $M$-PAM constellation spacing is 2, we have $\sqrt[n]{|det(\boldsymbol{G})|^2} = 4$, and the Poltyrev capacity condition for real lattices becomes $\sigma^2 < \frac{2}{\pi e}$, where for complex lattices it is $\sigma^2 < \frac{4}{\pi e}$. Interestingly, this is exactly the necessary condition that was developed above for the stack decoder to converge to the correct path. As this is a necessary but not sufficient condition, the stack decoder is not guaranteed to converge above capacity. Indeed, it is well known that sequential decoders can converge only above the cutoff SNR, which is approximately 1.7dB above capacity for the high SNR regime \cite{Forney_Ungerboeck}.

See \cite{sequential_lattice_isit} for another example of using the Fano metric for lattice decoding.

\subsection{Bidirectional Sequential Decoding}
\label{bidir_stack}
After developing the Fano metric for the stack (or Fano) algorithms, we shall now turn to develop a bidirectional decoding scheme for signal codes. It is well known that sequential decoding is sensitive to noise bursts \cite{Berlekamp_Jacobs}. In \cite{bidirectional}, a bidirectional decoding algorithm was proposed in order to reduce the complexity of decoding through a noise burst. Two stack decoders are working, where one works from the start of the block forward and the other moves from the end of the block backward. The algorithm stops when the two decoders meet at the same point. For a strong noise burst, each decoder will only have to face half the length of the burst. Assuming exponential complexity increase along the burst (since for strong noise, the entire tree has to be examined) the resulting complexity will be the square root of the complexity of a single decoder.

Note that in order to enable bidirectional decoding, the data must be partitioned to finite-length blocks, with known initial and final state. However, this is anyway the case for all the practical shaping algorithms that were presented in Section \ref{shaping}, as explained in Section \ref{termi}. The block length should be made as large as possible, such that the overhead of terminating the encoding in a known state will cause minimal degradation to information rate. However, increasing the block size introduces delay to the system. In addition, the probability to have two or more distinct strong noise bursts that appear in the same block increases. In such a case, each of the two decoders will have to face a strong noise burst alone, and bidirectional decoding will no longer be effective.

Unlike general lattice codes, bidirectional decoding is possible for signal codes due to the band-Toeplitz structure of the lattice generator matrix. However, decoding backward for signal codes is not straightforward, as reversing the time axis causes the minimum phase filter pattern to become maximum phase (i.e. all its zeros are outside the $Z$-plane unit circle). Extending the paths of the stack has an effect similar to filtering with an autoregressive filter with non-stable poles, resulting in choosing extension symbols that grow without bound. This can be easily solved by allpass filtering: if we filter the codeword (in the forward direction) with the allpass filter $A(z) = \frac{F^*(1/z^*)}{F(z)}$, then we have transformed the signal code to a code with a maximum-phase filter pattern. Decoding backward will now obey a stable recursion. Note that the allpass filtering does not change the power spectrum of the additive noise. See Section \ref{non_mp} for other applications of allpass filtering to signal codes. 

Bidirectional decoding is implemented using two stack decoders. Each stack decoder holds a stack of previously explored paths, where each path is assigned a score according to the Fano metric, as described above. Both decoders work simultaneously. At each step, the path with best score in the stack is extended to all its successors and then deleted from the stack. The successors then enter the stack. Before deletion, the deleted path is compared to all the paths of the stack of the other decoder to look for a merge. %For lattice decoding, the
%backward decoder performs QR decomposition for the matrix
%$\boldsymbol{H}$ with reversed column order.
% To go backward for lattice decoding, the order of columns of the matrix $\boldsymbol{H}$ is simply reversed, together with the order of elements in the vector $\boldsymbol{x}$ (See (\ref{Hx_n})). The QR factorization is performed again for the reversed $\boldsymbol{H}$, and then two separate decoders can work in parallel, where the order of examining the elements of $\boldsymbol{x}$ is reversed for both decoders.
A merge is declared when a path in the other decoder's stack is found with the same state at the same time point in the data block as the current decoder, i.e. last $L$ symbols of the forward decoder match the time-reversed last $L$ symbols of the backward decoder. In order to reduce the probability of false merge indications, a match of more than $L$ symbols can be used. However, as the number of bits in each extended constellation symbol $b_n$ is usually large (e.g. 17 bits for the real and imaginary parts for the example of Section \ref{NB_nature}), the probability of false indication is usually low enough for a match of $L$ symbols. %merge in a predefined number of symbols. 

A straightforward search for a merge will require a full pass on the whole stack every symbol. In order to avoid it, each stack entry can be assigned a hash value according to its last $L$ symbols. For each possible hash value, a linked list is maintained with all the stack entries that are assigned this value. Then, each decoder calculates the hash value that corresponds to its last $L$ symbols, and searches only the linked list of the other decoder that corresponds to this value, resulting in a much smaller search complexity. %False merges with large score are eliminated.

\section{Generalizations of the Basic Signal Coding Scheme} \label{extensions}%Extensions to 
%Non-Minimum Phase filters and IIR Filters}
\subsection{Non Minimum-Phase Filter Patterns} \label{non_mp}

Until now we have assumed that the filter pattern of the signal code is a minimum-phase filter. This assumption is essential for the recursive loops of the various shaping methods to be stable. We shall now show how to extend the concept of signal codes to non minimum-phase filters.

Denote a general invertible filter pattern by $F(z)=F_i(z)F_o(z)$, where $F_i(z)$ is a monic minimum phase filter and $F_o(z)$ is a monic maximum phase filter. We can deploy signal coding with the filter pattern $F_{MP}(z) = F_i(z)F_o^*(1/z^*)$, which is a minimum phase filter, and then apply an allpass filter $A(z) = F(z)/F_{MP}(z)$ to the encoded signal. The allpass filter does not change the signal power level or its power spectrum. Therefore, this scheme generates a lattice which is based on the filter pattern $F(z)$, which is not minimum-phase. As the recursive loops of the various shaping and encoding schemes work with the filter pattern $F_{MP}(z)$, which is minimum-phase, stability is ensured.

Note that both filters $F(z)$ and $F_{MP}(z)$ have the same frequency response magnitude and differ only in the frequency response phase. The following claim relates the error spectrum of the codes that relate to two filters with this property.

\begin{claim} \label{spect_claim}
Assume that $F_1(z)=1+\sum_i f_1(i) z^{-i}$ and $F_2(z) = 1+\sum_i f_2(i) z^{-i}$ are two filter patterns which are related by $F_2(z) = F_1(z) A(z)$, where $A(z)$ is an allpass filter. Then, every error-symbol sequence has the same Euclidean weight for the two signal codes that result from $F_1(z)$ and $F_2(z)$. In particular, the two codes have the same error spectrum.
\end{claim}
\begin{proof}
Assume that $e(n)$ is an error-symbol sequence %in the error spectrum of $F_1(z)$, 
with $Z$-transform $E(z)$. From (\ref{err_d_def}), its weight is $d_1^2(\underline{\boldsymbol{e}})= \sum_n \left|e(n)+ \sum_k f_1(k) e(n-k)\right|^2$. Using Parseval's rule, we have:
\begin{align*}
d_1^2(\underline{\boldsymbol{e}})=\frac{1}{2\pi} \int_0^{2\pi} \left|F_1(e^{jw})\right|^2 \left|E(e^{jw})\right|^2 dw.
\end{align*}
Calculating the weight of the same error-symbol sequence, but now for the filter pattern $F_2(z)$, we get:
\begin{align*}
d_2^2(\underline{\boldsymbol{e}})=\sum_n \left|e(n)+ \sum_k f_2(k) e(n-k)\right|^2 = 
\end{align*}
\begin{align*}
=\frac{1}{2\pi} \int_0^{2\pi} \left|F_2(e^{jw})\right|^2 \left|E(e^{jw})\right|^2 dw = 
\end{align*}
\begin{align*}
 =\frac{1}{2\pi} \int_0^{2\pi} \left|F_1(e^{jw})\right|^2 \left|A(e^{jw})\right|^2 \left|E(e^{jw})\right|^2 dw =
 \end{align*}
\begin{align*}
  =\frac{1}{2\pi} \int_0^{2\pi} \left|F_1(e^{jw})\right|^2 \left|E(e^{jw})\right|^2 dw = d_1^2(\underline{\boldsymbol{e}}).
\end{align*}
Therefore, every error-symbol sequence generates the same weight for both $F_1(z)$ and $F_2(z)$.
\end{proof}

Note that convolving the filter pattern of a signal code with an allpass filter is equivalent to multiplying a lattice generator matrix by an orthonormal matrix. Such a multiplication is equivalent to rotation and reflection of the lattice in Euclidean space, which do not change the error spectrum of the corresponding lattice code.

%For the AWGN channel, the error performance of the code is determined by its error spectrum. Therefore, 
Claim \ref{spect_claim} shows that non-minimum-phase filter patterns have no advantage over their minimum-phase equivalents when the AWGN is considered. However, in non AWGN channels, such as in fading channels and in impulse noise channels, mixed-phase channels may be advantageous since their impulse response may be longer, thus allowing better time-diversity.

\subsection{Auto-Regressive, Moving-Average (ARMA) Filter Patterns}
Thus far, we have described signal codes which employ FIR filter patterns, but the signal code concept can be easily extended to ARMA filter patterns. Suppose that we want to design a signal code with an ARMA filter pattern $F(z)=G(z)/H(z)$, where $G(z) = 1 +  \sum_{l=1}^L g_l z^{-l}$ and $H(z) = 1 +  \sum_{k=1}^K h_k z^{-k}$ are monic invertible minimum phase filters. The encoding operation will then be:
\begin{align}
	x(n) = b(n) +  \sum_{l=1}^L g_l b_{n-l} -  \sum_{k=1}^K h_k x_{n-k}
\end{align}

For Tomlinson-Harashima shaping, the shaping operation is:
\begin{align*} 
		b_n = a_n - 2M k_n.
\end{align*}
It can be easily seen that choosing
\begin{align*}
k_n = \left\lfloor \frac{1}{2M} \left( a_n +  \sum_{l=1}^L g_l b_{n-l} -  \sum_{k=1}^K h_k x_{n-k} \right) \right\rceil
\end{align*}
results in $\left|x(n)\right|\leq M$.
%while the extended symbols are generated by:
%
%		b(n) = a(n) - [  l=1..L g(l)b(n-l) -  k=1..K h(k)x(n-k) ]	
The other shaping methods of Section \ref{shaping} can be extended in a similar manner.
ARMA filter patterns can be particularly useful when signal coding is combined with channel pre-equalization, as described in the next subsection.
%The above signal encoders belongs to a broader class of signal encoding algorithms which consists of two steps: first the input data symbols are remapped into extended symbols that are drawn from the same lattice of the input constellation on the complex plane, and than the codeword is generated by filtering the extended symbols sequence with a monic stable filter, and possibly by another allpass filter. The remapping transformation must be invertible, so that it will not loose information, and its purpose is to optimize power of the codewords or to map to codewords that reside in a desired multi-dimensional hyper-shape. Consider now the N-dimensional Voronoi region of the input sequences of length N. In the case of M2-QAM inputs the input's Voronoi region is simply a multi-dimensional dimensional cube of side 1. In Appendix A we show that the flexible signal coder is the only encoder in this class that maps the input sequence to the Voronoi region that surrounds it. We also show that any encoder in the class which maps M2-QAM inputs to [-M,M)2N hypercube is identical up to permutation with the Tomlinson signal coder. 

\subsection{Combining Signal Coding with Pre-Equalization}
\begin{figure}
\centering
%\vspace {-0.3cm}
\includegraphics[width=3.5in]{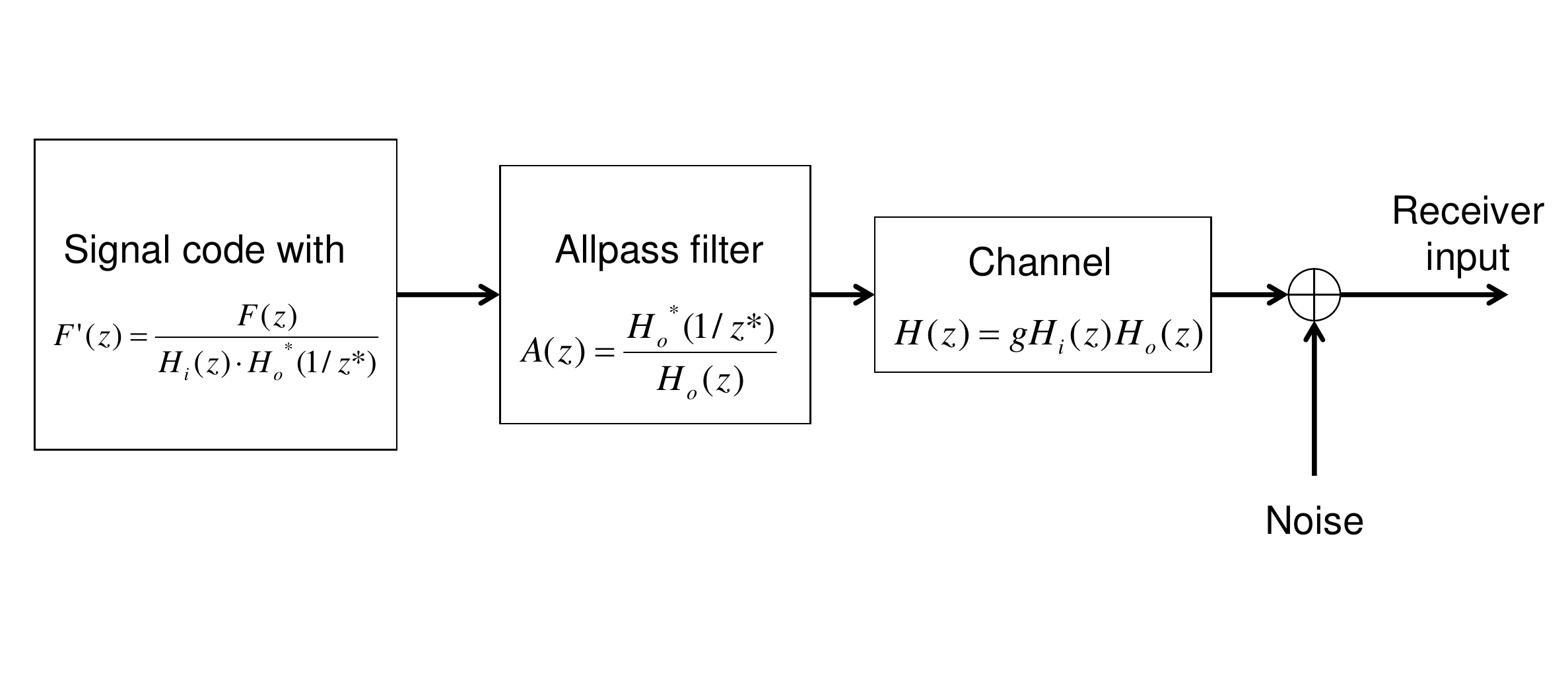} %{snr_e3.pdf}  %, width=2.5in
%\vspace {-0.5cm}
\caption{Combining signal coding with pre-equalization}
%\vspace{-0.7cm}
\label{pre_eq_fig}
\end{figure}

Assume that coding should be used for transmission through a communications channel which introduces inter-symbol interference (ISI). Signal coding can be seamlessly combined with channel pre-equalization, by designing the encoder's filter so that its convolution with the channel impulse response will be the desired signal code filter pattern, possibly up to a gain factor. However, this would work only if the channel is a minimum phase filter, since otherwise the encoder's filter is non-minimum phase and the recursive loops of its algorithms become unstable. In order to avoid this problem, we apply an all-pass filter to the transmitted signal, that converts the channel into a minimum phase system (this is a common procedure in equalization of digital communications channels \cite{Forney_Ungerboeck}). Let the channel be $H(z) = gH_i(z)H_o(z)$, where $g$ is a gain factor, $H_i(z)$ is a monic minimum phase filter, and $H_o(z)$ is a monic maximum phase filter. Assume further that $H(z)$ is stable and invertible. In order to transform the channel into its minimum phase equivalent, we apply the filter $A(z)=H_o^*(1/z^*)/H_o(z)$ to the channel input, transforming the combined channel $A(z)H(z)$ into a minimum phase system.  Since $A(z)$ is an allpass filter, i.e. $|A(e^{jw})|=1$, it does not affect the transmitted signal's power or power spectrum. We then apply the shaping and encoding operations using the monic minimum phase encoder filter $F'(z) = \frac{F(z)}{A(z) H_i(z) H_o(z)}$, where $F(z)$ is the desired signal code filter pattern. The resulting scheme is illustrated in Figure \ref{pre_eq_fig}. It can be easily seen that the linear system that relates $b(n)$ to the channel output, $F'(z)A(z)H(z)$, folds into the desired pattern $F(z)$, multiplied by the channel gain $g$. Therefore, the receiver can employ a detector that is optimized for an ideal (non-ISI) channel, and the error performance will be the same as in an ideal channel with a gain of $|g|$.  

\section{Simulation Results}
\label{sim_res}
We shall now demonstrate the performance of signal codes using simulations. All the simulations are for 6 bits per (complex) symbol (equivalent to uncoded 64-QAM). Unless otherwise stated, the simulations use the filter pattern $F(z)=(1+0.98e^{j0.09\pi}z^{-1})^3$ (the fourth filter pattern of Table \ref{dmin_table}), combined with Tomlinson-Harashima shaping (Section \ref{Tomli_shaping}). Data is framed to finite-length blocks, where block size is 2000 symbols. The total number of blocks that were simulated for each result is 20,000.

As explained in Section \ref{termi}, for a filter pattern of length $L+1$, the last $L$ values of $b_n$ should be transmitted at the end of each block. As shown in Section \ref{NB_nature}, 72 bits are required to store $L=3$ consecutive $b_n$'s for this specific filter pattern. In order to protect the $b_n$'s, 8-QAM modulation is used for their transmission. This way, the $b_n$'s are protected by approximately 9dB relative to uncoded 64-QAM. Since the gap to capacity for uncoded transmission at bit error rate (BER) of $10^{-6}$ is approximately 9dB \cite{Forney_Ungerboeck}, the uncoded $b_n$'s will be more protected than the coded data, so the error rate due to badly detected $b_n$'s is negligible.

Transmitting the 72 bits of the $b_n$'s using 8-QAM requires 24 symbols. Therefore, the actual information rate is not 6 bits/symbol but $6 \times \frac{2000}{2000+24} = 5.93$ bits/symbol.
To achieve unconstrained channel capacity of $6$ bits/symbol, the required SNR is $18$dB, where for $5.93$ bits/symbol, the required SNR is $17.8$dB. Therefore, data framing results in a loss of $0.2$dB. This loss is essentially an implementation loss and is not related to the coding properties of the signal code lattice. Note also that this implementation loss can be made negligible by increasing block length, or by using a more efficient coding scheme for transmitting the $b_n$ tail symbols. Since our main intention is to demonstrate the coding properties of the signal code lattice, and not the performance of the specific decoders, we shall ignore the framing loss and compare our results to channel capacity and cutoff rate for transmission of 6 bits/symbol. For the same reason, we shall use ideal path memories for the stack decoder (i.e. remember the full symbol path for each stack entry, and not use the more efficient methods of Section \ref{stack_dec}), in order to avoid the related implementation loss due to path memory truncation.

Since we use the Tomlinson-Harashima shaping scheme, the transmitted signal will be uniformly distributed. For $6$ bits/symbol under uniform input distribution constraint, channel capacity is at SNR of $19.1$dB, where the cutoff rate is at $20.9$dB. As the unconstrained capacity for $6$ bits/symbol is 18dB, the capacity loss due to the uniform distribution constraint is $1.1$dB. It can be seen that at these SNRs, the gap between the unconstrained capacity and the uniform distribution capacity has not reached yet its asymptotic value of $1.53$dB. The cutoff SNR is 1.8dB away from capacity, in accordance with the approximate 1.7dB gap mentioned in \cite{Forney_Ungerboeck}.
 %Since the slope of the channel capacity curve is approximately 3 dB/(bit/symbol) for high SNR, the loss of 0.07 bits/symbol is equivalent to approximately 0.2dB. Note that in all the performance curves of this section that show the gap to channel capacity and cutoff rate, this loss is ignored. The reason is that this loss can be made negligible by increasing block size, and since we mainly want to demonstrate the properties of the signal code lattice, and not the specific decoder.

\begin{figure}
\centering
%\vspace {-0.3cm}
\includegraphics[width=3.5in]{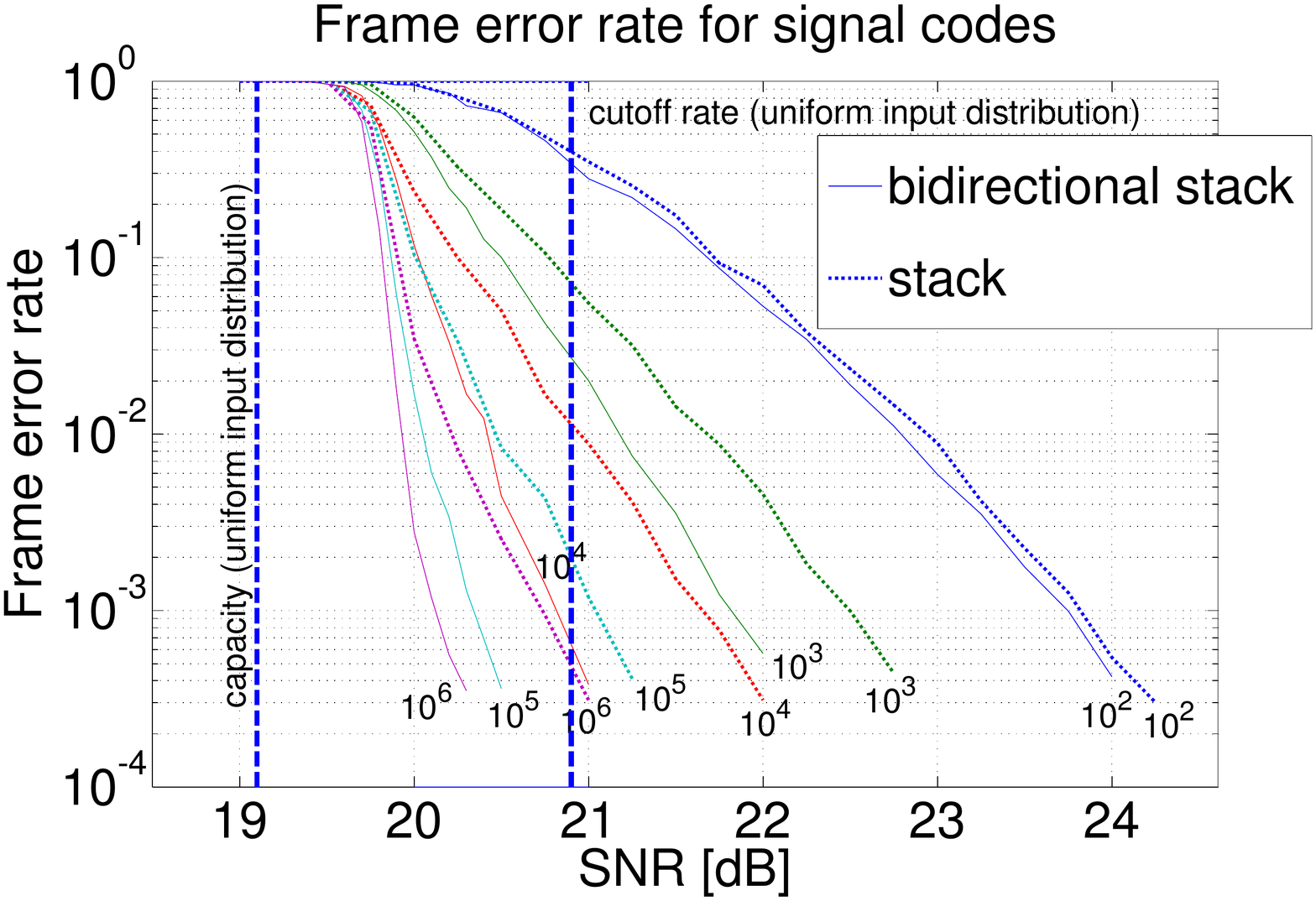} %{modified_ber} %{ber_uni_bidir.pdf}  %, width=2.5in
%\vspace {-0.5cm}
\caption{Frame error rate for stack and bidirectional stack decoding, for various maximal stack lengths. Each curve is labeled with the corresponding maximal stack length.}
%\vspace{-0.7cm}
\label{fer_sim_fig}
\end{figure}

\begin{figure}
\centering
%\vspace {-0.3cm}
\includegraphics[width=3.5in]{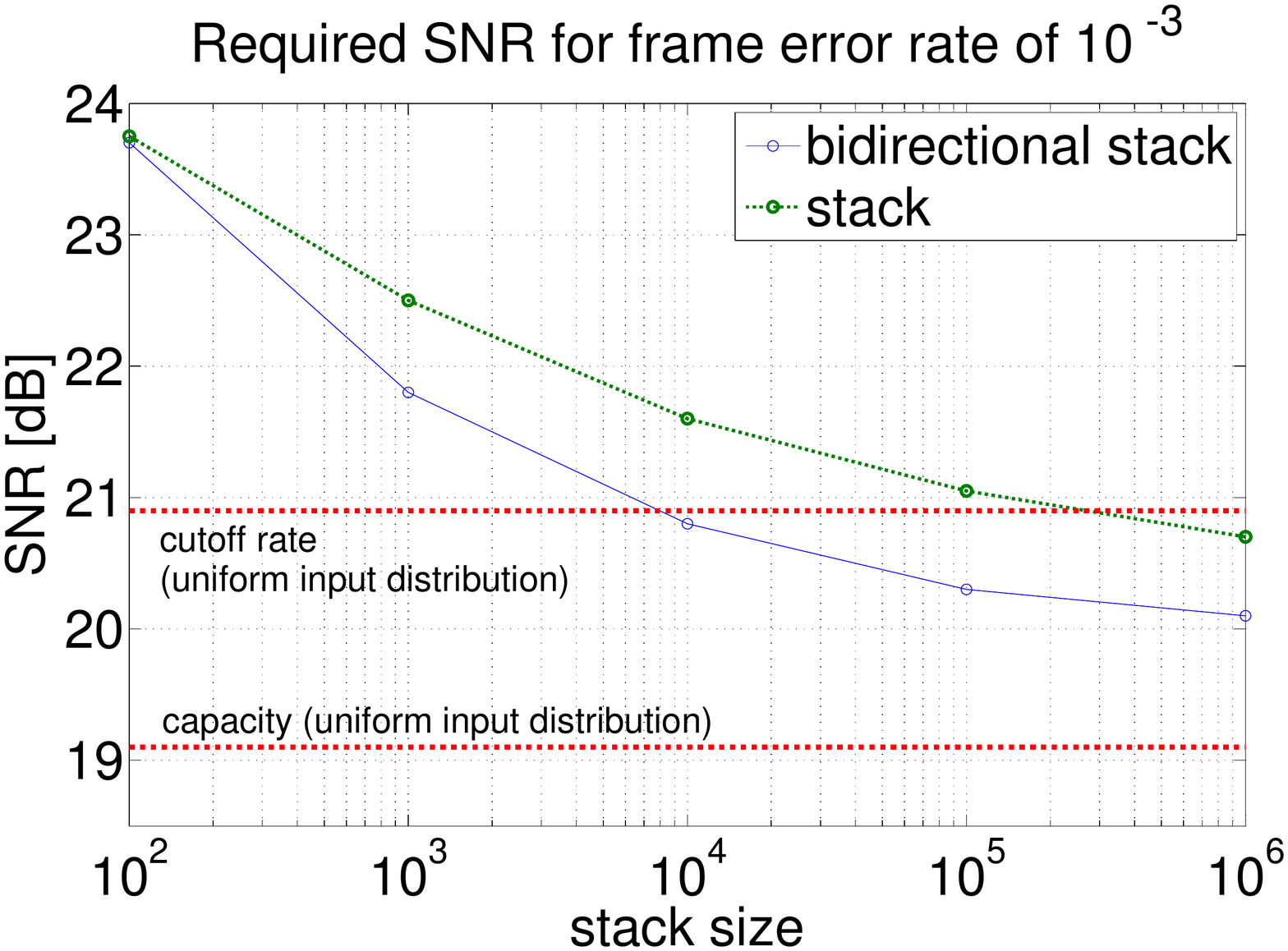} %{modified_comparison.pdf} %{snr_e3.pdf}  %, width=2.5in
%\vspace {-0.5cm}
\caption{required SNR to achieve frame error rate of $10^{-3}$ for the stack and bidirectional stack decoders}
%\vspace{-0.7cm}
\label{snr_sim_fig}
\end{figure}

%We shall demonstrate the performance of the stack and the bidirectional stack decoders. It comes out experimentally that best results were achieved by setting the bias parameter $B$ in the Fano metric (\ref{final_signal_metric}) to $1.2\sigma^2$, which slightly deviated from the calculated expression (\ref{B_def_complex}). For this bias, results were also better than setting the bias according to the non-approximated expression of (\ref{int_approx}).
%
Figure \ref{fer_sim_fig} shows the frame error rate (FER) vs. SNR using the stack and the bidirectional stack decoders. For each decoder, the FER is shown for various maximal stack lengths. The channel capacity and computational cutoff rate for $6$ bits/symbol with uniform channel input distribution are also shown in the figure. The same results are also presented in Figure \ref{snr_sim_fig}, where for each maximal stack length, the figure shows the required SNR for achieving frame error rate of $10^{-3}$. Note that this FER value is certainly a practical value for many applications, e.g. wireless networks. %For uncoded transmission with block size of 2,000 symbols at 6 bits/symbol, a frame error rate of $10^{-3}$ implies bit error rate (BER) of approximately $10^{-7}$.

It can be seen that increasing the maximal stack length improves the performance for both the stack and the bidirectional stack decoders. This can be explained as follows. When a noise burst is present, incorrect paths in the stack will temporarily have better score than the correct path. If the number of such incorrect paths exceeds the stack length, the correct path will be thrown out of the stack. This was defined in Section \ref{reduced_comp} above as a CPL event, which will result with a decoding error. Figures \ref{fer_sim_fig} and \ref{snr_sim_fig} show that for FER of $10^{-3}$ and stack length which is smaller than $10^6$, most of the errors result from CPL events and not from decoding to a wrong codeword that was closer to the observation in the Euclidean space, so increasing the stack length improves the FER.

It can be seen that with a very large stack length of $10^6$, and for frame error rate of $10^{-3}$, the stack decoder can work as close as $1.6$dB from channel capacity, which is 0.2dB beyond the channel cutoff rate. The bidirectional stack decoder can work as close as $1$dB from channel capacity, which is $0.8$dB beyond the cutoff rate. This is certainly a strong indication that the signal code lattice is indeed a capacity approaching lattice.
The fact that sequential decoders can work beyond the cutoff rate under these conditions should not be surprising: As explained in section \ref{reduced_comp}, for a fixed and finite frame error rate, sequential decoders can work beyond the cutoff rate with a finite (and probably large) computational complexity.

Turning to complexity, we shall now examine the computational and storage requirements of the decoders. The storage is determined by the maximal stack length, where the computational complexity can be defined by the average and maximal number of computations per symbol. For this purpose, a computation is defined as the processing of a single stack entry. The number of computations per a specific symbol is calculated by dividing the total number of computations for the block that contains this symbol, by the number of symbols in the block. The maximum and average over all the 20,000 blocks of each simulation are defined as the maximal and average number of computations per symbol, respectively. 

\begin{figure}
\centering
%\vspace {-0.3cm}
\includegraphics[width=3.5in]{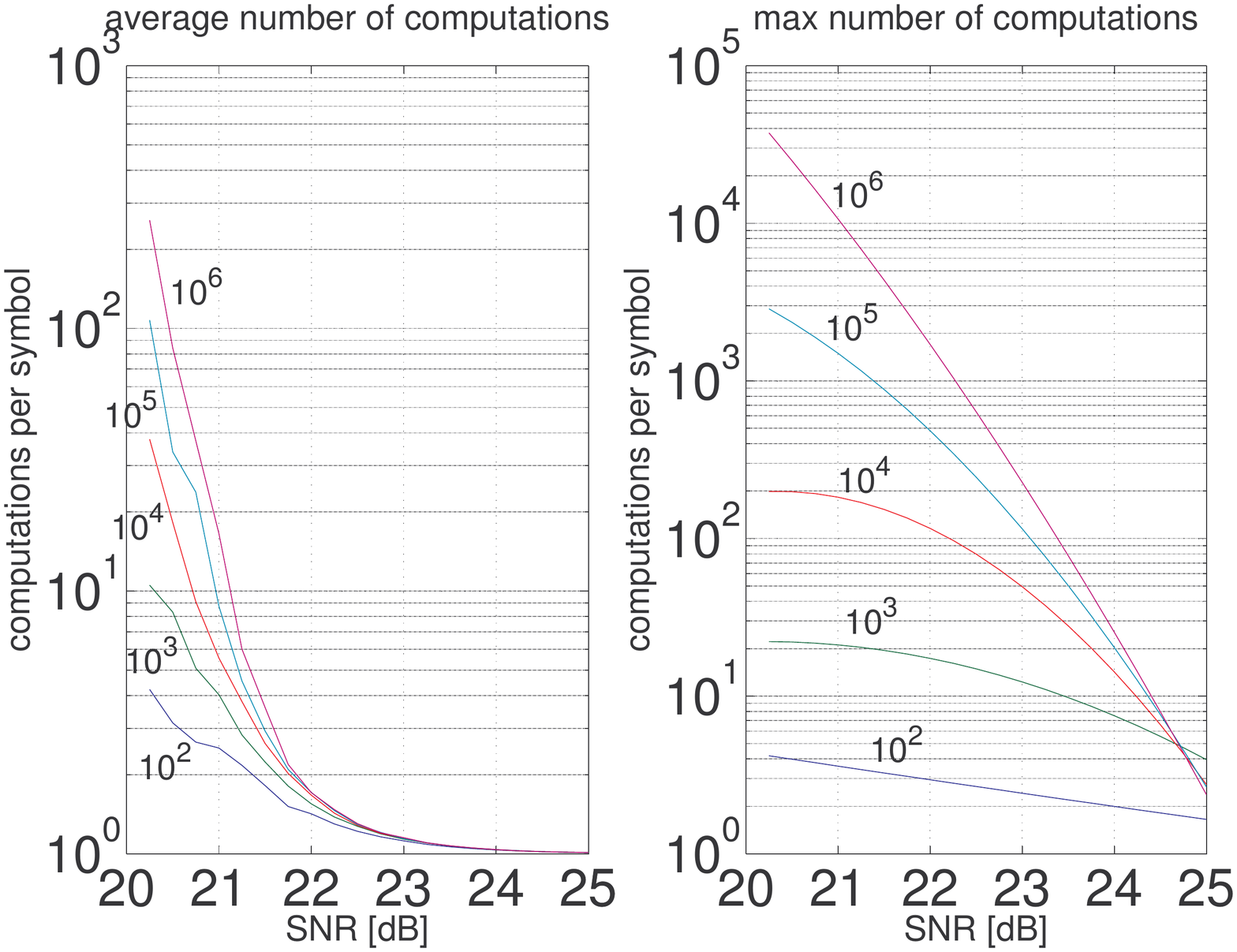}  %, width=2.5in
%\vspace {-0.5cm}
\caption{Average and maximal number of computations for the stack decoder. Each curve is labeled with the corresponding maximal stack length.}
%\vspace{-0.7cm}
\label{uni_comp_fig}
\end{figure}

\begin{figure}
\centering
%\vspace {-0.3cm}
\includegraphics[width=3.5in]{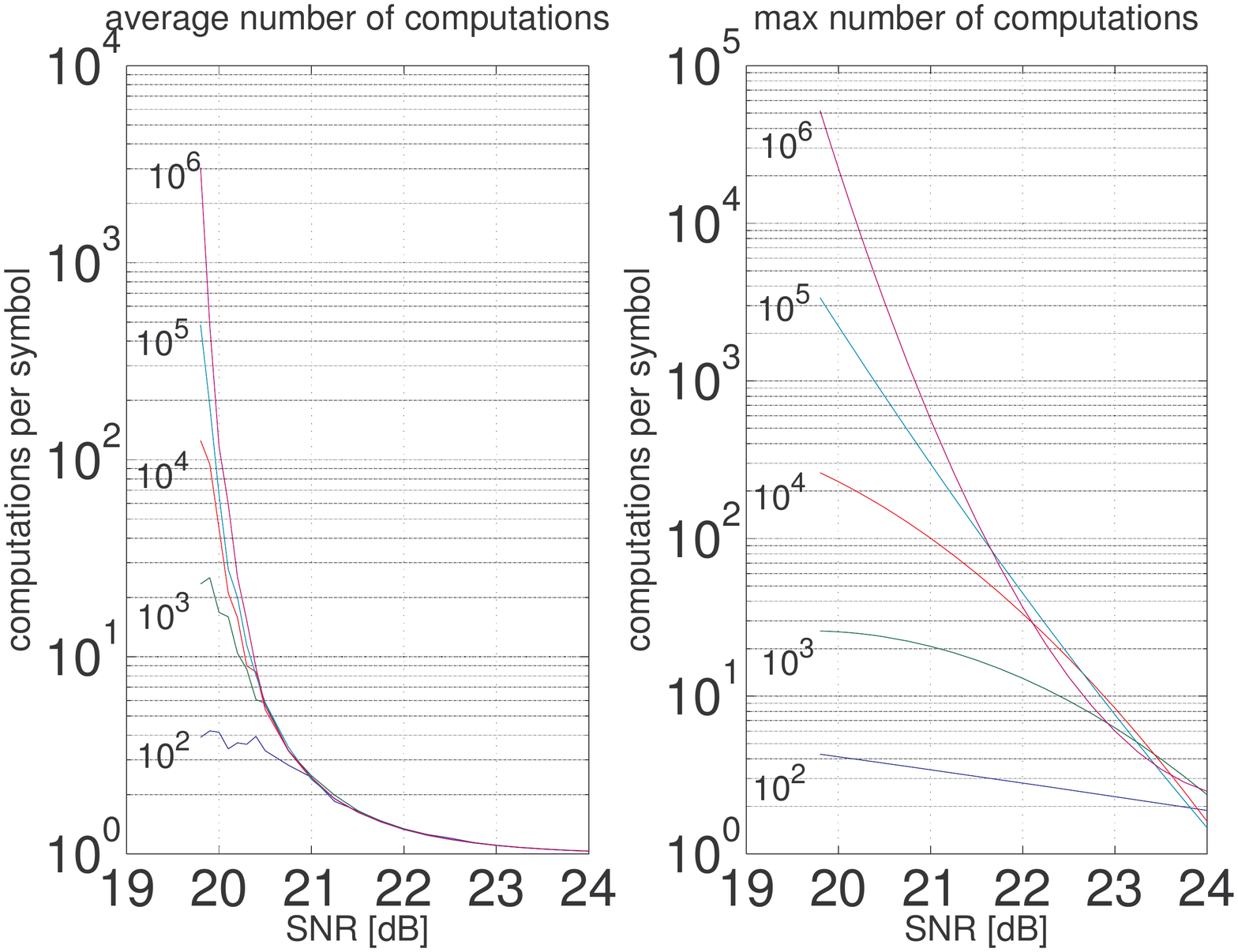}  %, width=2.5in
%\vspace {-0.5cm}
\caption{Average and maximal number of computations for the bidirectional stack decoder. Each curve is labeled with the corresponding maximal stack length.}
%\vspace{-0.7cm}
\label{bidir_comp_fig}
\end{figure}

Figure \ref{uni_comp_fig} shows the average and maximal number of computations for the stack decoder, where Figure \ref{bidir_comp_fig} shows it for the bidirectional stack decoder, for various maximal stack lengths. Combining the results from Figures \ref{snr_sim_fig} and \ref{bidir_comp_fig}, we can see that in order for the bidirectional stack algorithm to work at FER of $10^{-3}$ at $1$dB from capacity, we need a stack of size $10^6$. The average number of computations is 80 computations per symbol, which is certainly a practical number (similar to a 64-states Viterbi decoder, or to an LDPC code with average node degree of 10 that performs 8 iterations). However, the maximal number of computations per symbol is 15,000 - more than two orders of magnitude than the average. Therefore, such a decoder can be implemented with reasonable average complexity, but from time to time it will have large and unpredictable delays for the worst-case blocks.

A more practical scheme might be a bidirectional stack decoder with maximal stack length of $10^4$. FER of $10^{-3}$ can be achieved for SNR of 20.8dB (1.7dB from capacity). The average number of computations per symbol is only 3 computations/symbol, where the maximum is 120. This is certainly a practical scheme, where the effect of non-predictable decoding delays still exists, but it is much less severe.

Note that the the phenomenon of computational peaks also exists in modern iterative decoders, such as LDPC codes or turbo codes. For these codes, it is common to have a ``stopping criterion'', which stops decoding when the detected data is a valid codeword. In this case, most of the time the decoder performs a small number of iterations (e.g. 1-2), and from time to time it needs to perform more iterations (e.g. 8-16). This will result in non-uniform processing complexity. However, the ``peak-to-average'' of the number of computations is still significantly larger for the proposed sequential decoders.

\begin{figure}
\centering
%\vspace {-0.3cm}
\includegraphics[width=3.5in]{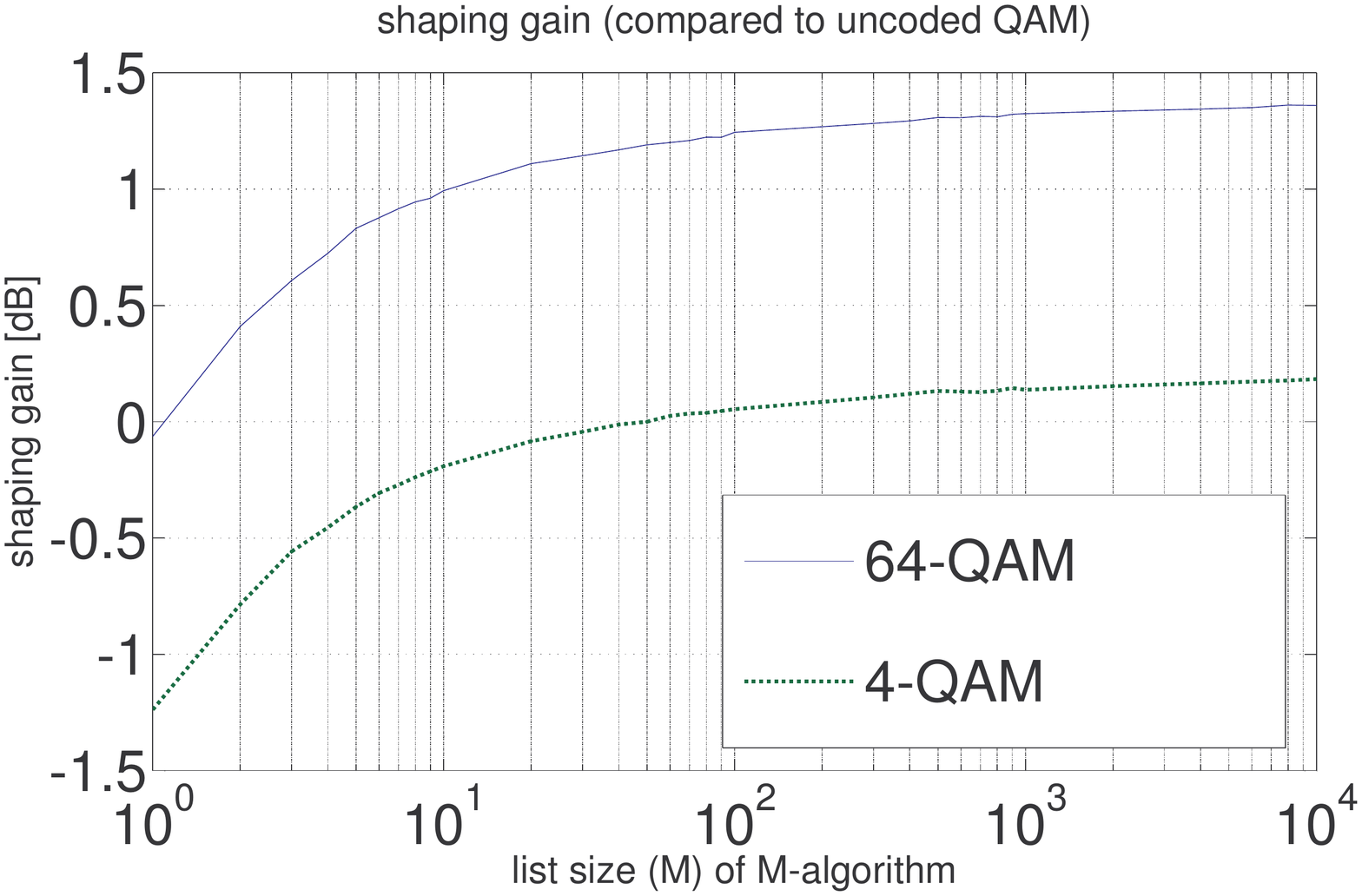}  %, width=2.5in
%\vspace {-0.5cm}
\caption{Nested lattice shaping gain for 64-QAM and 4-QAM constellations.}
%\vspace{-0.7cm}
\label{shaping_gain_fig}
\end{figure}

All the results so far were presented for Tomlinson-Harashima shaping. With this scheme, the codeword elements are uniformly distributed, so no shaping gain can be attained relative to uncoded QAM. However, such shaping gain can be achieved using nested lattice shaping, as explained in Section \ref{nested}. %We shall now demonstrate the performance of nested lattice shaping (Section \ref{nested}). Unfortunately, we could not demonstrate simple simulation results, %(As in Figure \ref{})
%since it comes out that for nested lattice shaping the stack and bidirectional stack decoders can not work much beyond the cutoff rate. The reason is that when Tomlinson-Harashima shaping is used, the decoder can dilute the search tree by ignoring states where the resulting channel symbol of the examined codeword is outside the allowed range. This can be easily done since the channel symbols are uniformly distributed in a finite interval. %(See Section \ref{}). 
%However, when nested lattice shaping is used, the channel symbols are not bounded in value - only their average energy is minimized. Therefore, it is not simple to dilute the search tree and the decoder's computational complexity is higher. Therefore, a better decoder has to be found in order to utilize the full shaping gain of nested lattice shaping and still work close to capacity.
%
%However, 
In order to understand the potential shaping gain of nested lattice shaping, 
Figure \ref{shaping_gain_fig} shows the average energy of the nested lattice shaper output, compared to the energy of uncoded QAM symbols. Nested lattice shaping was implemented using the M-algorithm \cite{Aulin_M}, as described in Section \ref{nested}. %. This algorithm works sequentially on the input symbols of the block, and at each stage stores the $M$ sequences that were found so far with minimum energy. For each symbol, each of the $M$ entries is extended with all ``reasonable'' values for $k_n$ (those that do not generate an immediate large energy penalty). All the extended sequences are sorted, and the $M$ which result in smallest energy are kept as input to the next stage. The value of $M$ determines both the storage and the computational complexity of the shaper. 
For $M=1$, nested lattice shaping reduces to Tomlinson-Harashima shaping.
As explained in Section \ref{Tomli_shaping}, the Tomlinson-Harashima scheme has an energy penalty of $M^2/(M^2-1)$ relative to uncoded $M^2$-QAM. For 64-QAM, this penalty is 0.07dB, where for 4-QAM it is 1.25dB. This explains the values of both curves of Figure \ref{shaping_gain_fig} for $M=1$. As $M$ increases, the shaping gain increases, and reaches 1.4dB for 64-QAM, which is close to the theoretical limit. For 4-QAM, the energy penalty of the Tomlinson-Harashima scheme is completely compensated, with additional gain of 0.2dB. Note that most of the shaping gain can be achieved with a practical $M$ value of 100 (1.25dB gain for 64-QAM and 0dB for 4-QAM).

Note that the computational complexity of the stack and the bidirectional stack decoders is much larger when nested lattice shaping is used, compared to the case where Tomlinson-Harashima shaping is used. The reason is that for the Tomlinson-Harashima scheme, ``x-range testing'' can be used to dilute the stack, as described in Section \ref{stack_dec}. Therefore, in addition to the increased complexity at the encoder side, nested lattice shaping has also a complexity penalty at the decoder side. This is a topic for further study.

\section {Summary}
A novel lattice coding scheme was introduced. Signal codes are based on projecting the conventional PAM/QAM signal points into filtered lattices that have better distance spectra. Error analysis and simulation results indicate that the signal code lattice is capacity approaching. Low complexity schemes based on Signal codes were demonstrated to attain the cutoff rate of the AWGN channel, where higher complexity schemes were demonstrated to work approximately 1dB from channel capacity.

\section*{Acknowledgment}
Support and discussions with Ehud Weinstein and Dave Forney are gratefully acknowledged. 
%% optional entry into table of contents (if used)
%%\addcontentsline{toc}{section}{Acknowledgment}
%Support and interesting discussions with Ehud Weinstein are gratefully acknowledged.

%\appendices
%\section{Volume of Signal Codes Basic Cell} \label{cell_volume}
%The submatrix formed from the first $N-L$ rows of $\boldsymbol{G}$ is a square lower triangular matrix with '$1$'s on the diagonal, so its determinant equals $1$. Now, the volume of the basic cell of a lattice with square generator matrix $\boldsymbol{G}$ is $det(\boldsymbol{G})$, where for a general $m \times n$ generator matrix $\boldsymbol{G}$ with $m \geq n$ the volume is $\sqrt{det(\boldsymbol{G}'\boldsymbol{G})}$.

\appendices
\section{Detailed Description of the Algorithm for Calculating the Error Spectrum} \label{err_spect_alg_app}

\begin{figure}
\centering
%\vspace {-0.3cm}
\includegraphics[width=3.5in]{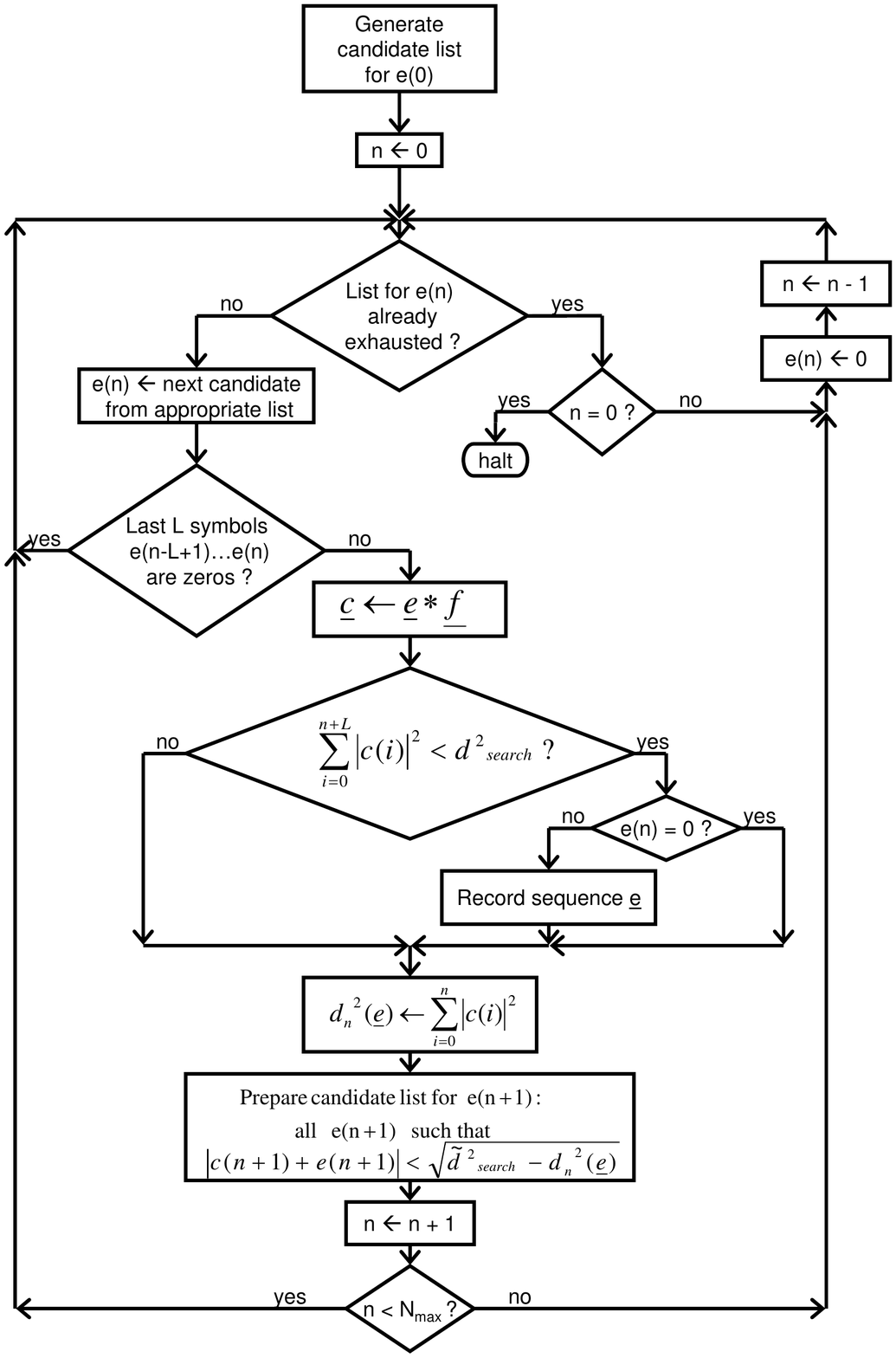}  %, width=2.5in [height=1.7in]
%\vspace {-0.5cm}
\caption{Algorithm for calculating the error spectrum of signal codes}
%\vspace{-0.7cm}
\label{err_spect_flow}
\end{figure}

Consider a signal code with a given filter pattern $F(z)$. The filter's impulse response sequence $f(0), f(1),...,f(L)$ will be denoted by $\underline{f}$. We shall now present an algorithm that finds all the error sequences whose Euclidean weight is below a given $d_{Search}^2$, where the length of the appropriate error-symbol sequence $\underline{\boldsymbol{e}}$ is smaller than $N_{max}$ symbols.
The flowchart of the algorithm is shown in Figure \ref{err_spect_flow}. Basically, it develops a tree of all possible error sequences, and truncates tree branches as soon as it can identify that all the error events on them will have distances above $d_{Search}^2$. The tree is searched in a Depth First Search (DFS) manner, which can be easily implemented using recursion techniques.

The basic step of the algorithm is as follows. Assume that we have built so far an error-symbol sequence of $n+1$ symbols $e(0), e(1), ..., e(n)$. Denote this sequence by $\underline{e}$. We want to extend this sequence with another symbol $e(n+1)$ such that the Euclidean weight of the resulting error sequence (and its possible extensions) can still be lower than $d_{Search}^2$. During the extension process, we would like to record all the error-symbol sequences for which the Euclidean weight of the resulting error sequence is actually smaller than $d_{Search}^2$. 

We start by calculating the convolution of $\underline{e}$ with the filter pattern, $\underline{c} = \underline{e} * \underline{f}$ (the length of $\underline{c}$ is $n+L$). Then, if the Euclidean norm of $\underline{c}$ is smaller than $d_{Search}^2$, $\underline{e}$ is recorded as an error event, after verifying that the last symbol $e(n)$ is nonzero (otherwise, each sequence will be recorded multiple times with zero padding). Then, we calculate the Euclidean norm of the first $n+1$ elements of $\underline{c}$, $d^2_n(\underline{e})=\sum_{i=0}^n \left|c(i)\right|^2$. This term will be part of the Euclidean weight of any error sequence that starts with $e(0), e(1), ..., e(n)$, and since $F(z)$ is a casual filter, it is independent of $e(n+k)$ for all $k>0$. %The reason is that when $\underline{e}$ is extended with an extra symbol $e(n+1)$, the first $n+1$ elements of the convolution of the resulting sequence with $\underline{f}$ will not change and will still equal $c(0), c(1),...,c(n)$. 
%The new error symbol $e(n+1)$ will only affect the next convolution element. 
As the filter $F(z)$ is monic, the next convolution element equals $c(n+1)+e(n+1)$. A necessary condition for the resulting error sequence to have Euclidean weight less than $d_{Search}^2$ is therefore:
\begin{align} \label{ext_cond}
d^2_n(\underline{e}) + \left|c(n+1)+e(n+1)\right|^2 < d_{Search}^2.
\end{align} 
A candidate list is built for $e(n+1)$ which includes all the values of $e(n+1)$ that satisfy (\ref{ext_cond}).

The computational complexity of the algorithm can be further improved by using a modified bound $\tilde{d}_{Search}^2$ in (\ref{ext_cond}), where $\tilde{d}_{Search}^2 = d_{Search}^2 - 4\ \left|f_L\right|^2$. The term $4\left|f_L\right|^2$ is a lower bound on the Euclidean weight of the convolution tail, since this will be the weight of the last tail symbol in case that we already reached the last nonzero symbol of the error sequence and its magnitude is the smallest possible symbol magnitude (i.e. $2$). As the test of (\ref{ext_cond}) uses the Euclidean weight of the error sequence \emph{without} encountering the convolution tail, and as the weight of the convolution tail is lower bounded by this term, we can truncate  branches whose weight has exceeded $\tilde{d}_{Search}^2$ instead of $d_{Search}^2$, thus reducing the tree search complexity.

The candidate list for the first error symbol $e(0)$ is built in a different manner than for the other error symbols. For $e(0)$, the candidate list contains all possible complex integers with even real and imaginary parts whose squared magnitude is smaller than $\tilde{d}_{Search}^2$. In order to make the algorithm more efficient, specific properties of the signal code lattice can be used to dilute this list. First, the convolution operation is shift invariant, so every error event will appear in the error spectrum with all its possible shifted versions. Therefore, we can eliminate the zero symbol from the candidate list for $e(0)$, such that shifted versions of the same error event will not be encountered. Also, using symmetry, if a complex integer $c$ is in the candidate list for $e(0)$, we can dilute from the list the values $-c$, $jc$ and $-jc$, where $j=\sqrt{-1}$, since these will result in the same error events, up to multiplication by the constants $-1$, $j$, $-j$, respectively.

We can now describe the flow of the algorithm, as shown in Figure \ref{err_spect_flow}. The algorithm starts by building a candidate list for the first error symbol $e(0)$.
Starting with $n=0$, the algorithm passes at each tree node over the candidate list elements for the next symbol $e(n)$, one by one. For each element, it first checks if the resulting $\underline{\boldsymbol{e}}$ sequence ends with $L$ zeros, in which case it skips to the next element in the list. This is a non-interesting error event as it is simply the concatenation of two non-overlapping error events. %$\underline{\boldsymbol{e}}$ is convolved with the filter pattern, and the score variable is updated with the Euclidean weight of the result (excluding the convolution tail, which may be varied by future symbols in the sequence). If the Euclidean weight of the whole convolution result (including the tail) is smaller than $d_{Search}^2$, the sequence is recorded as an error event, after verifying that the last $e_n$ is nonzero (otherwise, each sequence will be recorded multiple times with zero padding). 
Then, a candidate list is constructed for the next symbol $e(n+1)$, while appropriate error-symbol sequences are recorded, using the basic step of the algorithm, as explained above. If the sequence length has not yet exceeded the maximum allowed length $N_{max}$, the algorithm repeats this procedure for the next tree node.
When the candidate list for $e(n)$ is exhausted, the algorithm goes back one step in the tree and continues with the candidate list that was previously prepared for $e(n-1)$. When the candidate list for $e(0)$ is finally exhausted, the algorithm terminates.

Note that instead of calculating the convolution $\underline{c} = \underline{e} * \underline{f}$ and the partial weight $d^2_n(\underline{e})$ at each tree node, a simple recursive update can be applied to the results of the calculations at the parent tree node, thus reducing the computational complexity. Also, instead of actually storing the candidate lists for the error symbols, the appropriate candidate can be calculated at each node where only an index needs to be stored.

Note also that if only the minimal distance of the code needs to be found, the computational complexity of the algorithm can be reduced by dynamically updating $d_{Search}^2$: it can be initialized to infinity, and whenever an error sequence with Euclidean weight smaller than $d_{Search}^2$ is recorded, $d_{Search}^2$ is updated to the weight of this sequence.

We finally note that the complexity of the error spectrum search algorithm of Figure \ref{err_spect_flow} can be further improved by using a ``backward-forward'' approach. With this approach, the algorithm first builds a \emph{tails-database}, which stores all the possible tail sequences whose Euclidean distance %is smaller than 
%finds all the error sequences such that the Euclidean distance of their ``tail''  %$\sum_{n>L} \left|e_n+ \sum_{k=1}^L f_k e_{n-k}\right|^2$ 
is lower than $d_{Tail}^2$. %, and stores them in a \emph{tails-database}. 
This can be done by applying the algorithm of Figure \ref{err_spect_flow} backwards in time. The algorithm then develops the error tree forward in time, but the condition for keeping an error sequence in the tree is that either its Euclidean weight % $\sum_{n=1}^N \left|e_n+ \sum_{k=1}^L f_k e_{n-k}\right|^2$ 
is smaller than $d_{Search}^2-d_{Tail}^2$, or that the last $L-1$ elements of the error sequence coincide with the first $L-1$ elements of an error sequence from the tails database, in which case their concatenation may yield an error event whose distance is below $d_{Search}^2$. This way, the effective search radius of the forward search is $d_{Search}^2-d_{Tail}^2$ instead of $d_{Search}^2$, which may result in significant complexity reduction even for relatively small values of $d_{Tail}^2$.
%The search over the tails database can be implemented with a computational complexity of the base-2 logarithm of the data-base size, or even with lower complexity, using hash tables (similarly to finding a merge in the bidirectrional stack algorithm, as described in Section \ref{bidir_stack}). %We also note that this backwards-forward concept can be applied to analyze the distance spectrum of any code which is a finite state machine (e.g. a trellis code) in a channel which is a finite state machine (e.g. ISI channel). 

\section{The Error Spectrum of the Cartesian lattice} \label{Cartesian_calc}
We shall now find the error spectrum of a simple lattice - the Cartesian lattice, whose generator matrix is the identity matrix, and can be interpreted as a signal code with $F(z)=1$. The error spectrum of such a lattice will include sequences whose elements are complex integers with even real and imaginary parts. Consider the set of infinite sequences of this form whose Euclidean weight is finite, and whose elements are restricted to be nonzeros.
It can be easily seen that the Euclidean weight of such sequences must be an integral multiple of 4. %: $\sum_{i=1}^\infty q_i^2 = 4k$, $k \in \mathbb{Z}$. 
Denote by $a(k)$ the number of such sequences whose Euclidean weight equals $4k$. Denote by $b(k)$ the number of such sequences whose Euclidean weight equals $4k$ and are further restricted to contain a single nonzero symbol. 
\begin{claim}
$a(k)$ and $b(k)$ are related by the following recursion:
\begin{align} \label{cart_recur}
a(k) = b(k) + \sum_{i=1}^{k-1} a(i) b(k-i)
\end{align}
\end{claim}
\begin{proof}
Adding a single nonzero symbol increases the Euclidean weight of a sequence by at least 4. Therefore, if we remove a single symbol from a sequence with weight $4k$, then the resulting weight will be at most $4(k-1)$. As a result, every sequence of weight $4k$ is a concatenation of a sequence with weight smaller or equal to $4(k-1)$ and a single symbol, and the recursion (\ref{cart_recur}) follows.
\end{proof}

The values of $b(n)$ are simple to calculate manually. It can be easily seen that the first 13 values are \{4, 4, 0, 4, 8, 0, 0, 4, 4, 8, 0, 0, 8\}. Starting with $a(1)=4$ and using (\ref{cart_recur}), we get that the first 10 values of $a(n)$ are \{4,  20,  96,  468,  2280,  11104,  54080,  263380,  1282724,  6247176\}. It can be seen that the error spectrum increases exponentially with the Euclidean weight. Note that this is a lower bound on the error spectrum, as we have ignored sequences which may contain zero symbols.

\section{Derivation of the Fano Metric for Signal Codes} \label{Fano_app}
Consider the following transmission model through a discrete, memoryless channel whose input and output are complex numbers in $\mathbb{C}$. The transmission uses a variable length code whose codewords $\left\{\boldsymbol{x}_1, \boldsymbol{x}_2,...,\boldsymbol{x}_M\right\}$ have lengths $\left\{n_1, n_2,...,n_M\right\}$, respectively. Let $x_{m,i}$ denote the $i$-th coordinate of $\boldsymbol{x}_M$. Let $S_i=\cup_m\left\{x_{m,i}\right\}$ be the set of all possible complex values for the coordinate $x_{m,i}$. Let $|S_i|$ denote the cardinal number of $S_i$, %, and assume that $|S_i|<\infty$.
and let $N \geq \max_m(n_m)$. To each codeword $\boldsymbol{x}_m=\left[x_{m,0}x_{m,1}\cdots x_{m,n_m-1}\right]$, having probability $P_m$, a random tail $\boldsymbol{t}_m=\left[t_{m,n_m}\cdots t_{m,N-1}\right]$ is appended, where $t_{m,j} \in S_j$, producing the word $\boldsymbol{z}=\left[z_0 z_1 \cdots z_{N-1}\right] = \left[x_{m,0}x_{m,1}\cdots x_{m,n_m-1} t_{m,n_m}\cdots t_{m,N-1}\right]$, which is sent over the channel. It is assumed that $t_{m,j}$ are independent of each other and of $\boldsymbol{x}_m$, for $n_m \leq j \leq N-1$. Let $p_j(\cdot)$ denote the probability distribution function of $t_{m,j}$. As explained in \cite{Fano_metric}, this decoding problem is essentially the same problem of choosing the best path in each step of the stack algorithm, where the stack contains paths of different lengths.

By independence, $Pr(\boldsymbol{t}_m|\boldsymbol{x}_m) = Pr(\boldsymbol{t}_m) = \prod_{k=n_m}^{N-1} p_k(t_{m,k})$. Let $\boldsymbol{y}=(y_0,y_1,y_2,...,y_{N-1})\in \mathbb{C}^N$ denote the received word. The joint probability distribution of appending a tail $\boldsymbol{t}_m$ to a codeword $\boldsymbol{x}_m$ and receiving $\boldsymbol{y}$ is:
\begin{align}
\boldsymbol{f}(\boldsymbol{x}_m,\boldsymbol{t}_m,\boldsymbol{y}) = P_m Pr(\boldsymbol{t}_m|\boldsymbol{x}_m)\boldsymbol{f}(\boldsymbol{y}|\boldsymbol{x}_m, \boldsymbol{t}_m) = 
\end{align}
\begin{align*}
= P_m Pr(\boldsymbol{t}_m)\boldsymbol{f}(\boldsymbol{y}|\boldsymbol{x}_m, \boldsymbol{t}_m) =
\end{align*}
\begin{align*}
= Pm \prod_{k=n_m}^{N-1}p_k(t_k)\prod_{k=0}^{n_m-1} f(y_k|x_{m,k})\prod_{k=n_m}^{N-1}f(y_k|t_{m,k}).
\end{align*}
Summing over all random tails gives the marginal distribution
\begin{align}
\boldsymbol{f}(\boldsymbol{x}_m, \boldsymbol{y}) = P_m \prod_{k=0}^{n_m-1} f(y_k|x_{m,k}) \prod_{k=n_m}^{N-1}f_k(y_k),
\end{align}
where:
\begin{align} \label{y_i_def}
f_k(y_k) = \sum_{w \in S_k} f(y_k|w)p_k(w).
\end{align}
Given $\boldsymbol{y}$, the maximum a posteriori decoding rule is to choose $\boldsymbol{x}_m$ which maximizes $P_r(\boldsymbol{x}_m|\boldsymbol{y})$. Equivalently,
\begin{align*}
\boldsymbol{f}(\boldsymbol{x}_m, \boldsymbol{y})/\prod_{i=0}^{N-1} f_k(y_k)
\end{align*}
can be maximized, as the denominator is independent of $\boldsymbol{x}_m$. Taking logarithms, the final statistic to be maximized by the optimum decoder is 
\begin{align} \label{Fano_score}
\boldsymbol{L}(\boldsymbol{x}_m, \boldsymbol{y}) = \sum_{i=0}^{n_m-1}\left[\log\left(\frac{f(y_i|x_{m,i})}{f_i(y_i)}\right) + \frac{1}{n_m} \log(P_m)\right]
\end{align}

Interestingly, the statistic for each codeword depends only on that portion of the received word $\boldsymbol{y}$ having the same length as the codeword.

We can now derive the Fano metric for the decoding of signal codes in the AWGN channel with noise variance $\sigma^2$. For simplicity, we shall start with real valued signal codes, and then extend the results to the complex case.
Assume that the data symbols $\{a_n\}$ are $M$-PAM symbols. There are $M$ possible symbols, so the a-priori probability of a codeword of length $n_m$ is:
\begin{align} \label{P_m_term}
P_m = \frac{1}{M^{n_m}} = M^{-n_m}
\end{align}

The numerator of the left term inside the sum of (\ref{Fano_score}) is:
\begin{align} \label{Gauss_term}
f(y_i|x_{m,i}) = \frac{1}{\sqrt{2\pi\sigma^2}}e^{-(y_i-x_{m,i})^2/2\sigma^2}
\end{align}

In order to calculate the denominator, we shall assume that Tomlinson-Harashima shaping is used. In this case, the set $S_i$, as defined above, is a finite set of values, uniformly spread in the interval (-$M$, $M$]. We shall assume that $|S_i|$ is large, such that we can approximate the sum of (\ref{y_i_def}) by an integral:
\begin{align} \label{int_approx}
f_i(y_i) = \sum_{w \in S_i} f(y_i|w)p_i(w) \approx 
\end{align}
\begin{align*}
\approx \int_{-M}^M \frac{1}{\sqrt{2\pi\sigma^2}}e^{-(y_i-w)^2/2\sigma^2} \frac{1}{2M} dw =
\end{align*}
\begin{align*}
=  \frac{1}{2M} \int_{\frac{-M-y_i}{\sigma}}^{\frac{M-y_i}{\sigma}} \frac{1}{\sqrt{2\pi}}e^{-z^2/2}  dz =
\end{align*}
\begin{align*}
= \frac{1}{2M} \left[Q\left(\frac{-M-y_i}{\sigma}\right) - Q\left(\frac{M-y_i}{\sigma}\right)\right]
\end{align*}
where $Q(x) \buildrel \Delta \over = \frac{1}{\sqrt{2\pi}} \int_x^{\infty} e^{-z^2/2}  dz$.
Note that the integral of (\ref{int_approx}) is a convolution between a rectangular pulse and a Gaussian. Assuming $\sigma^2 << M$ (high SNR), the Gaussian is much narrower than the rectangular pulse, so the convolution result can be approximated by a rectangular pulse with height $\frac{1}{2M}$, except for values of $y_i$ that are relatively close to the edges of the pulse at $\pm M$. We can then simply approximate (\ref{int_approx}) by the constant $\frac{1}{2M}$, assuming that the probability of $y_i$ being near the edges can be neglected. %If we now take into account that $y_i$ is actually complex, and further assume that due to the Tomlinson-Harashima shaping operation the real and imaginary parts of $y_i$ are independent, 
We then get:
\begin{align} \label{const_approx}
f_i(y_i) \approx \frac{1}{2M}.
\end{align}
Substituting (\ref{P_m_term}), (\ref{Gauss_term}) and (\ref{const_approx}) in (\ref{Fano_score}) and organizing terms, we finally get:
\begin{align} \label{final_signal_metric}
\boldsymbol{L}(\boldsymbol{x}_m, \boldsymbol{y}) = \sum_{i=0}^{n_m-1}\left[-(y_i-x_{m,i})^2 + B\right]
\end{align}
where
\begin{align} \label{B_def}
B \buildrel \Delta \over = \sigma^2 \cdot \log \frac{2}{\pi \sigma^2}
\end{align}

The extension of these results to complex signal codes with $M^2$-QAM input constellation and complex noise variance of $\sigma^2$ is straightforward. Instead of (\ref{P_m_term}), (\ref{Gauss_term}) and (\ref{const_approx}) we have $P_m = M^{-2n_m}$, $f(y_i|x_{m,i}) = \frac{1}{\pi\sigma^2}e^{-(y_i-x_{m,i})^2/\sigma^2}$ and $f_i(y_i) \approx \frac{1}{4M^2}$, respectively (where we have assumed that the Tomlinson-Harashima precoding causes the real and imaginary parts to be independent of each other). Substituting in (\ref{Fano_score}), we get again expression (\ref{final_signal_metric}), where now we have:
\begin{align} \label{B_def_complex}
B \buildrel \Delta \over = \sigma^2 \cdot \log \frac{4}{\pi \sigma^2}
\end{align}

\end{document}